\documentclass[twocolumn, twocolappendix]{aastex631}

\usepackage{amsmath}
\usepackage{bm}

\usepackage[capitalise]{cleveref} 
\crefname{section}{Section}{Sections}
\Crefname{section}{Section}{Sections}
\crefname{equation}{Equation}{Equations}
\crefname{figure}{Figure}{Figures}

\makeatletter
\usepackage{etoolbox}
\patchcmd\H@refstepcounter{\protected@edef}{\protected@xdef}{}{}
\makeatother

\defcitealias{PB2022reanalysis}{PB22}
\defcitealias{PB2020BB}{PB20}
\defcitealias{class2020}{P20}

\usepackage{multirow}

\shorttitle{Measuring Atmospheric circular polarization}
\shortauthors{Fujino, T. \& S. Takakura et al.}

\begin{document}

\title{
A measurement of atmospheric circular polarization with POLARBEAR
}

\author[0000-0002-1211-7850]{Takuro Fujino}
\affiliation{Graduate School of Engineering Science \& Faculty of Engineering, Yokohama National University, Kanagawa 240-8501, Japan}

\author[0000-0001-9461-7519]{Satoru Takakura}
\affiliation{International Center for Quantum-field Measurement Systems for Studies of the Universe and Particles (QUP), High Energy Accelerator Research Organization (KEK), Tsukuba, Ibaraki 305-0801, Japan}
\affiliation{Department of Physics, The University of Tokyo, Tokyo 113-0033, Japan}

\author[0009-0005-4168-2858]{Shahed Shayan Arani}
\affiliation{Department of Physics, University of California, San Diego, La Jolla, CA 92093, USA}

\author[0000-0002-1623-5651]{Darcy Barron}
\affiliation{Department of Physics and Astronomy, University of New Mexico, Albuquerque, New Mexico 87131, USA}

\author[0000-0002-8211-1630]{Carlo Baccigalupi}
\affiliation{The International School for Advanced Studies (SISSA), Via Bonomea 265, I-34136 Trieste, Italy}
\affiliation{The National Institute for Nuclear Physics (INFN), Via Valerio 2, I-34127 Trieste, Italy}

\author[0000-0002-3266-857X]{Yuji Chinone}
\affiliation{International Center for Quantum-field Measurement Systems for Studies of the Universe and Particles (QUP), High Energy Accelerator Research Organization (KEK), Tsukuba, Ibaraki 305-0801, Japan}
\affiliation{Kavli Institute for the Physics and Mathematics of the Universe (WPI), UTIAS, The University of Tokyo, Kashiwa, Chiba 277-8583, Japan}

\author[0000-0002-1419-0031]{Josquin Errard}
\affiliation{Universit\'{e} Paris Cité, CNRS, Astroparticule et Cosmologie, F-75013 Paris, France}

\author[0000-0002-3255-4695]{Giulio Fabbian}
\affiliation{Kavli Institute for Cosmology Cambridge, Madingley Road, Cambridge CB3 0HA, UK}
\affiliation{Institute of Astronomy, Madingley Road, Cambridge CB3 0HA, UK}

\author[0000-0001-7438-5896]{Chang Feng}
\affiliation{Department of Astronomy, University of Science and Technology of China, Hefei 230026, China}
\affiliation{School of Astronomy and Space Science, University of Science and Technology of China, Hefei 230026, China}

\author[0000-0003-2606-9340]{Nils W. Halverson}
\affiliation{Department of Astrophysical \& Planetary Sciences, University of Colorado Boulder, UCB391, Boulder, Colorado 80309, USA}

\author[0000-0003-1443-1082]{Masaya Hasegawa}
\affiliation{Institute of Particle and Nuclear Studies (IPNS), High Energy Accelerator Research Organization (KEK), Tsukuba, Ibaraki 305-0801, Japan}

\author[0000-0001-6830-8309]{Masashi Hazumi}
\affiliation{International Center for Quantum-field Measurement Systems for Studies of the Universe and Particles (QUP), High Energy Accelerator Research Organization (KEK), Tsukuba, Ibaraki 305-0801, Japan}
\affiliation{Institute of Particle and Nuclear Studies (IPNS), High Energy Accelerator Research Organization (KEK), Tsukuba, Ibaraki 305-0801, Japan}
\affiliation{Kavli Institute for the Physics and Mathematics of the Universe (WPI), UTIAS, The University of Tokyo, Kashiwa, Chiba 277-8583, Japan}
\affiliation{Japan Aerospace Exploration Agency (JAXA), Institute of Space and Astronautical Science (ISAS), Sagamihara, Kanagawa 252-5210, Japan}
\affiliation{School of High Energy Accelerator Science, The Graduate University for Advanced Studies, SOKENDAI, Kanagawa 240-0193, Japan}

\author[0000-0001-5893-7697]{Oliver Jeong}
\affiliation{Department of Physics, University of California, Berkeley, Berkeley, California 94720, USA}

\author[0000-0003-3917-086X]{Daisuke Kaneko}
\affiliation{International Center for Quantum-field Measurement Systems for Studies of the Universe and Particles (QUP), High Energy Accelerator Research Organization (KEK), Tsukuba, Ibaraki 305-0801, Japan}

\author[0000-0003-3118-5514]{Brian Keating}
\affiliation{Department of Physics, University of California, San Diego, La Jolla, CA 92093, USA}

\author[0009-0004-9631-2451]{Akito Kusaka}
\affiliation{Department of Physics, The University of Tokyo, Tokyo 113-0033, Japan}
\affiliation{Physics Division, Lawrence Berkeley National Laboratory, Berkeley, CA 94720, USA}
\affiliation{Research Center for the Early Universe, School of Science, The University of Tokyo, Tokyo 113-0033, Japan}
\affiliation{Kavli Institute for the Physics and Mathematics of the Universe (WPI), UTIAS, The University of Tokyo, Kashiwa, Chiba 277-8583, Japan}

\author[0000-0003-3106-3218]{Adrian Lee}
\affiliation{Department of Physics, University of California, Berkeley, Berkeley, California 94720, USA}
\affiliation{Physics Division, Lawrence Berkeley National Laboratory, Berkeley, CA 94720, USA}

\author[0000-0001-9002-0686]{Tomotake Matsumura}
\affiliation{Kavli Institute for the Physics and Mathematics of the Universe (WPI), UTIAS, The University of Tokyo, Kashiwa, Chiba 277-8583, Japan}

\author[0000-0001-7868-0841]{Lucio Piccirillo}
\affiliation{Jodrell Bank Centre for Astrophysics, School of Physics and Astronomy, University of Manchester, Manchester, M13 9PL, UK}

\author[0000-0003-2226-9169]{Christian L. Reichardt}
\affiliation{School of Physics, University of Melbourne, Parkville, VIC 3010, Australia}

\author[0000-0001-5667-8118]{Kana Sakaguri}
\affiliation{Department of Physics, The University of Tokyo, Tokyo 113-0033, Japan}

\author[0000-0001-6830-1537]{Praween Siritanasak}
\affiliation{National Astronomical Research Institute of Thailand, Chiangmai, Thailand 50180}

\author[0000-0003-0221-2130]{Kyohei Yamada}
\affiliation{Department of Physics, The University of Tokyo, Tokyo 113-0033, Japan}

\correspondingauthor{Takuro Fujino}
\email{fugino-takuro-yk@ynu.jp}

\correspondingauthor{Satoru Takakura}
\email{t.satoru99@gmail.com}

%
%
%
%



\begin{abstract}
At millimeter wavelengths, the atmospheric emission is circularly polarized owing to the Zeeman splitting of molecular oxygen by the Earth's magnetic field.
We report a measurement of the signal in the 150\,GHz band using 3 years of observational data with the \textsc{Polarbear} project.
Non-idealities of a continuously rotating half-wave plate (HWP) partially convert circularly polarized light to linearly polarized light.
While \textsc{Polarbear} detectors are sensitive to linear polarization, this effect makes them sensitive to circular polarization.
Although this was not the intended use, we utilized this conversion to measure circular polarization.
We reconstruct the azimuthal gradient of the circular polarization signal and measure its dependency from the scanning direction and the detector bandpass.
We compare the signal with a simulation based on atmospheric emission theory, the detector bandpass, and the HWP leakage spectrum model.
We find the ratio of the observed azimuthal slope to the simulated slope is $0.92 \pm 0.01\rm{(stat)} \pm 0.07\rm{(sys)}$.
This ratio corresponds to a brightness temperature of $3.8\,\mathrm{m K}$ at the effective band center of $121.8\,\mathrm{GHz}$ and bandwidth of $3.5\,\mathrm{GHz}$ estimated from representative detector bandpass and the spectrum of Zeeman emission.
This result validates our understanding of the instrument and reinforces the feasibility of measuring the circular polarization using the imperfection of continuously rotating HWP.
Continuously rotating HWP is popular in ongoing and future cosmic microwave background experiments to modulate the polarized signal.
This work shows a method for signal extraction and leakage subtraction that can help measuring circular polarization in such experiments.
\end{abstract}



\section{Introduction} \label{sec:introduction}
The cosmic microwave background (CMB) provides a powerful way to probe the history of the universe.
From the observation of the CMB temperature anisotropy, we can accurately determine the parameters of the standard model of cosmology, the so-called $\Lambda$ Cold Dark Matter ($\Lambda$CDM) model \citep{planck2018vi}.
Moreover, the CMB is linearly polarized.
CMB linear polarization observations can be used for further tests of the $\Lambda$CDM model, such as the value of the optical depth at reionization, and to test the theory of cosmological inflation.
\citep{SO2019, S4_science_book, litebird2023}.

Circular polarization of the CMB is another component we can explore.
The standard cosmological model predicts that the CMB should not be circularly polarized at the last scattering surface.
However, certain mechanisms can generate circular polarization in the CMB photons during their propagation from the last scattering to an observer today.
Examples are Faraday conversion by the magnetic fields of galaxy clusters \citep{circ_galaxy},
Faraday conversion by relativistic plasma remnants of Population III stars 
 \citep{circ_popIII},
scattering by the cosmic neutrino background (C$\nu$B) 
 \citep{circ_CnB},
and photon--photon scattering \citep{circ_photon_photon}.
There also are predictions of the circular polarization of the CMB arising from extension of the standard model of cosmology and particle physics,
such as pseudoscalar fields \citep{circ_pseudo_field}
and Lorentz violation \citep{circ_lorentz}.

Furthermore, the atmospheric emission is circularly polarized \citep{atmcp_lenoir1968, Keating1998ApJ, Hanany2003, Spinelli2011, class2020}.
Under the presence of the Earth's magnetic field, molecular oxygen in the atmosphere undergoes Zeeman splitting.
In the frequency range of most CMB experiments, the dominant emission lines are in the range of 50 to 70\,GHz and at 118.8\,GHz.
This emission is circularly polarized.
Thus, this emission becomes a foreground contamination for the observation of CMB circular polarization from the ground.
Galactic and other foregrounds, such as synchrotron emission from the Galaxy, are summarized in \citet{king2016circular}.

The Zeeman emission is generally considered in the context of remote sensing of the temperature of the mesosphere,
which observes the atmospheric emission at frequencies around the resonance lines \citep{Meeks1963, Guzman2015}.
Even at frequencies away from the resonance lines, 
circularly polarized emission can be measured using polarization-sensitive microwave telescopes such as those used to observe the CMB.
The CLASS experiment observed the circular polarization signal in the 40\,GHz band \citep{class_clvv2020,class2020}.

Although modern CMB instruments are optimized to be sensitive to linear polarization, and not insensitive to circular polarization, there are several ways to obtain circular polarization data from them.
One method is to convert circular polarization to linear polarization using a half-wave plate (HWP) exploiting frequency-dependent non-idealities \citep{circ_SPIDER_result}.
A second method is to use a variable-delay polarization modulator as in the CLASS experiment.
A third method is to use a quarter-wave plate, which is an optical device that converts circular polarization to linear polarization.
Finally, a fourth method is to use coherent receivers like those of the Very Large Array \citep{thompson1980very} and the Atacama Large Millimeter/Submillimeter Array \citep{brown2004alma}.
Recently, SPIDER and CLASS have provided upper limits on the degree of circular polarization of CMB \citep{circ_SPIDER_result,class_clvv2020,class_clvv2024}.
SPIDER employed HWPs rotating in steps of $22.5$ degrees, and CLASS used a variable-delay polarization modulator.
In this study, we analyze \textsc{Polarbear} data collected using a continuously rotating HWP.

This paper reports the observation of atmospheric circularly polarized signal using \textsc{Polarbear} data.
\textsc{Polarbear} is a CMB experiment that began in January 2012.
From May 2014 to 
January 2017,
\textsc{Polarbear} performed large-angular-scale observations using a HWP, which was continuously rotated at 2\,Hz
\citep{PB2022reanalysis}.
We measure the circular polarization making use of the non-idealities of the HWP and study the atmospheric circular polarization signal analyzing its azimuthal and spectral dependencies.

In \Cref{sec:model}, we explain
the models used in this paper for the atmospheric circularly polarized emission, the HWP, and the detector response.
In \Cref{sec:data}, we summarize the \textsc{Polarbear} instrument and data used in this analysis.
The analysis method is explained in \Cref{sec:analysis}.
We report the results in \Cref{sec:results}.
Finally, we conclude the paper in \Cref{sec:conclusion}.

\section{Model} \label{sec:model}
\subsection{Atmospheric circular polarization}\label{sec:atm model}

The atmosphere is one of the main sources of contamination in ground-based CMB experiments.
Atmospheric molecules absorb part of the astronomical signals and emit thermal radiation around their resonance frequencies.
Around CMB observation bands, molecular oxygen has strong resonances around 60 and 119\,GHz, 
and water vapor has strong resonances around 22, 183, and 325\,GHz~\citep{am}.
The bandpass filters of detectors are designed to avoid these lines (see \cref{sec:data}).

These atmospheric emissions are mostly unpolarized \citep[see, e.g.][]{errard2015modeling}.
However, there is some polarized radiation.
One example of such polarized radiation is the horizontal linear polarization from tropospheric ice clouds, which has been detected by \textsc{Polarbear} in the 150\,GHz band~\citep{Takakura2019ApJ}, by CLASS in the 40, 90, 150, and 220\,GHz bands~\citep{class_cloudpol}, and by SPT-3G in the 95, 150, and 220\,GHz bands~\citep{spt_cloudpol}.
Another contaminant is circular polarization due to the Zeeman effect of molecular oxygen created by the Earth's magnetic field.
This signal has been measured by the CLASS experiment in the 40\,GHz band~\citep[\citetalias{class2020} hereafter]{class2020}.
In this work, we focus on the last example of circularly polarized radiation.

We adopt the theoretical model of atmospheric circular polarization described in \citetalias{class2020}.
Here, we briefly explain its properties.

The oxygen molecule is the only abundant molecule in the atmosphere with a non-zero magnetic moment.
The magnetic moment results from the two electrons in the highest energy state coupling with parallel spin. 
Owing to Zeeman splitting caused by the Earth's magnetic field, the energy level at quantum number $J$ is split into $2J+1$ equally spaced levels.
Here, $J$ is the rotational quantum number and $M$ ($|M| \leq J$) is the magnetic quantum number.
According to the selection rules, only transitions with $\Delta J = \pm 1$ ($J = N \to J = N \pm 1$) are permitted.
We refer to these transitions as the $N^+$ transition for $J=N+1$ and $N^-$ transition for $J=N-1$ respectively.
In this situation, the selection rules permit the transition of $\Delta M = 0, \pm 1$ at each $N$.
The emission from the $\Delta M = 0$ transition creates linear polarization,
whereas the emission from the $\Delta M = \pm 1$ transition creates both linear and circular polarization.
The Zeeman splitting breaks emission balance between right-handed and left-handed circular polarization from the $\Delta M = \pm 1$ transitions,
and circular polarization signal appears above and below the resonance frequencies.
In \textsc{Polarbear} case, the amplitude of expected circular polarization signal is approximately three orders of magnitude larger than the amplitude of the linear polarization signal owing to the Zeeman splitting effect.
In the simulation reported in this paper, we look at the values of $N$ from 1 to 37.
In the case of the oxygen molecule, the frequencies of the resonance lines are 118.8\,GHz for $N^-=1^-$ and between 50 to 70\,GHz for other values of $N$.
Thus, the circular polarization signal appears above and below the 118.8\,GHz line and from 50 to 70\,GHz with overlapping lines.
The strength of these circular polarization signals depends on the angle between the line of sight and the Earth's magnetic field\footnote{The azimuth and elevation of the Earth's magnetic field vector at the \textsc{Polarbear} site are $-5.6^\circ$ and $20.9^\circ$, respectively.}.
The strength reaches a maximum when the angle is $0\degr$ or $180\degr$ and zero when the angle is $90^\circ$. In fact, the Stokes $V$ signal $\propto \cos \theta_{\rm{az}}$, where $\theta_{\rm{az}}$ is the azimuthal angle between the line of sight and the Earth's magnetic field.
Moreover, this amplitude depends on the elevation angle of the pointing direction.
Regarding the elevation dependence, the signal amplitude depends on the angle between the line of sight and the Earth's magnetic field, as well as the change in the atmosphere optical depth along the line of sight.
The signal is stronger at low elevation because of the contribution of both effects.
\begin{figure}
\centering
\includegraphics[width=\columnwidth]{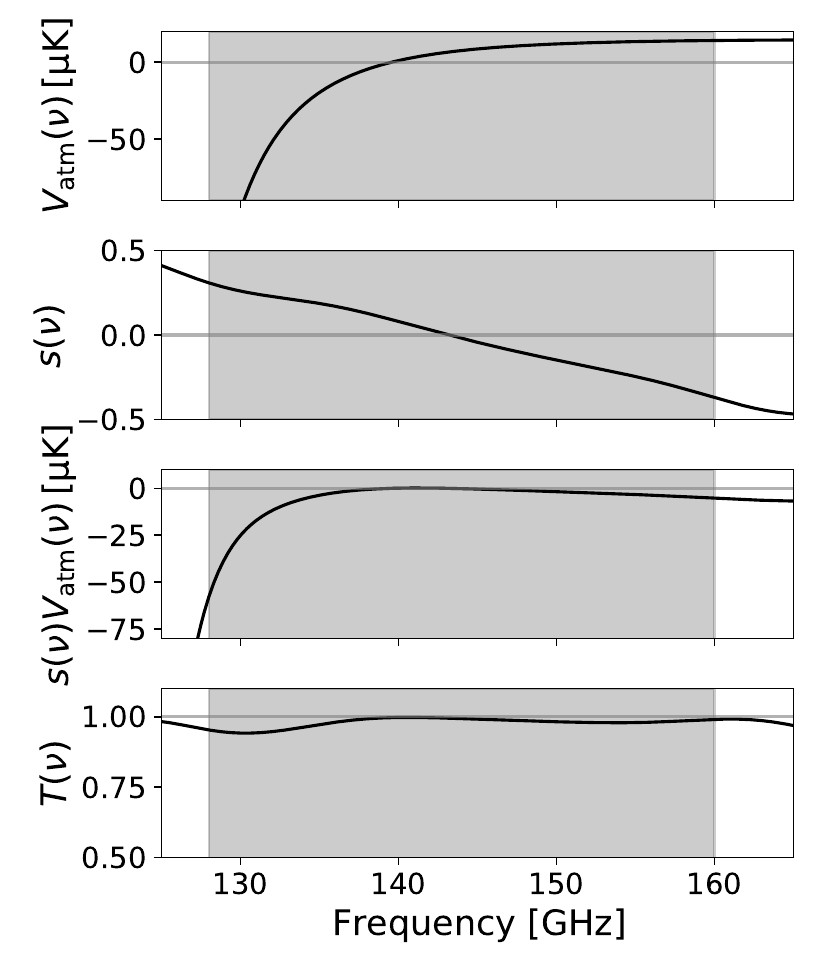}
\caption{(Top) Brightness temperature spectrum of the atmospheric circular polarization signal.
(Upper Middle) Spectrum of the conversion factor from $V$ to $U$ of the HWP.
(Lower middle) Combination of the two effects.
(Bottom) The transmittance of the HWP.
The shaded region shows the averaged detector bandpass shown in \cref{tab:bandpass}.
\label{fig:model_freq_dep}}
\end{figure}

\subsection{Half-wave plate} \label{sec:HWP}
\textsc{Polarbear} adopts a rotating HWP to modulate the linear polarization signal\footnote{A quarter-wave plate is ideal for observing circular polarization while it halves the sensitivity to linear polarization.}.
An ideal HWP is not supposed to convert circular polarization to linear polarization.
In practice, however, the ideal condition is satisfied only at a specific frequency, and non-zero conversion occurs at other frequencies in the frequency band of the instrument.
Using this non-ideality of the HWP, an instrument sensitive to linear polarization becomes sensitive to circular polarization \citep{circ_SPIDER_result}.

The optical properties of the \textsc{Polarbear} HWP were measured in the laboratory before deployment in the field~\citep{Fujino2023}.
The conversion factor from the circular polarization (Stokes $V$) to linear polarization (Stokes $U$) is expressed as a parameter $s$ in the Mueller matrix \citep{HWPmodel_SPIDER}:
\begin{equation}
    \begin{pmatrix}
        I_\mathrm{out}(\nu) \\ Q_\mathrm{out}(\nu) \\ U_\mathrm{out}(\nu) \\ V_\mathrm{out}(\nu)
    \end{pmatrix} =
    \begin{pmatrix}
        T(\nu) & \rho(\nu) & 0 & 0 \\
        \rho(\nu) & T(\nu) & 0 & 0 \\
        0 & 0 & c(\nu) & -s(\nu) \\
        0 & 0 & s(\nu) & c(\nu)
    \end{pmatrix}
    \begin{pmatrix}
        I_\mathrm{in}(\nu) \\ Q_\mathrm{in}(\nu) \\ U_\mathrm{in}(\nu) \\ V_\mathrm{in}(\nu)
    \end{pmatrix},
\end{equation}
where $T(\nu)$, $\rho(\nu)$, and $c(\nu)$ are the transmittance, differential transmittance between the two HWP axes, and polarization efficiency, respectively.
We calculate the spectrum of $s (\nu)$ following \cite{HWP_modeling}.
The estimated spectrum of $s(\nu)$ is shown in \cref{fig:model_freq_dep}.
$s(\nu)$ is zero around the band center and becomes positive or negative at lower and higher frequencies, respectively.

\subsection{Simulations}
\label{sec:simulation}

We simulate the expected signal using the models above.
The effective circular polarization signal $\overline{sV}$ is the product of the source spectrum $V(\nu)$ (in brightness temperature units) and $s(\nu)$ averaged over the detector bandpass function $W(\nu)$:
\begin{equation}
\label{eq:band averaged sV}
\overline{sV}
= \frac{\int s(\nu) V(\nu)W(\nu) d\nu}{\int W(\nu) 
 T(\nu) d\nu}\;.
\end{equation}
In this calculation, the source spectrum is divided by the HWP transmittance $T(\nu)$
because the instrument is calibrated using thermal sources and the HWP transmittance is not included in $W(\nu)$.
The $T(\nu)$ spectrum is shown in \cref{fig:model_freq_dep}.

We calculate the band-averaged $sV$ values at each of the seven detector wafers (see \cref{sec:instrument}).
Because of the strong frequency dependence of the atmospheric signal,
the instrument is sensitive to the average $V$ at the lower frequency edge of the bandpass.
The bandpass function of \textsc{Polarbear} detectors was measured at the site \cite[see also \cref{sec:data}]{Matsuda2019}.
The detector bandpass within each wafer is very uniform, whereas there are slight differences between wafers.
The integration range is twice the bandwidth above and below the center frequency.
This range is roughly from $85\, \rm{GHz}$ to $205\, \rm{GHz}$.
In the following sections, values without explicit reference to $\nu$ indicate band-average values.

\begin{figure}
    \centering
    \includegraphics[width=\linewidth]{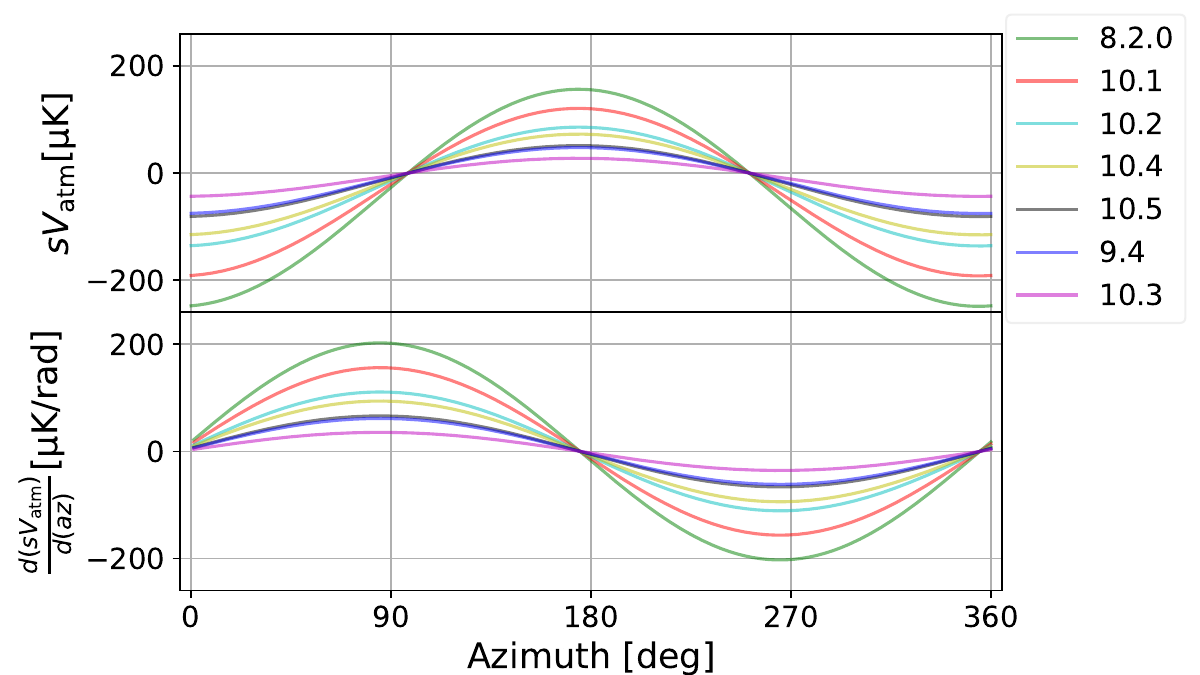}
    \caption{Azimuthal dependence of the $sV$ signal and its derivative with respect to the azimuthal angle.
    The elevation assumed in this simulation is 30\,degrees. 
    Each color represents a detector wafer.
    The legend on the right side of the plot is arranged by wafer, starting with those that have the lowest lower edge of the bandpass.}
    \label{fig:sV sim}
\end{figure}
\cref{fig:sV sim} shows the simulated $sV$ signal for each wafer.
The amplitude of the cosine curve increases as the lower-edge of the bandpass decreases.
\cref{fig:sV sim} also shows the derivative of the $sV$ signal with respect to the azimuthal angle.
The absolute value of the derivative has a maximum value of $\sim 200\,\mathrm{\mu K/rad}$ around azimuthal angles of 90 and 270 degrees.

We also estimate the effective band center and bandwidth using the spectra above.
To properly handle the Zeeman spectrum, we assume the Zeeman spectrum to be symmetrical around the resonance line and define an effective bandpass $W'(\nu)$ that folds back at the resonance frequency $\nu_0 = 118.8\,\mathrm{GHz}$:
\begin{equation}
    W'(\nu) =
    \begin{cases}
        W(\nu) s(\nu) - W(2\nu_0 - \nu) s(2\nu_0 - \nu) & (\nu \geq \nu_0) \\
        0 & (\nu < \nu_0)
    \end{cases},
\end{equation}
and calculate the effective band center $\nu_c$ as:
\begin{equation}
    \int^{\nu_c} |W'(\nu) V(\nu)| d\nu = 0.5 \int |W'(\nu) V(\nu)| d\nu.
\end{equation}
Then we calculate $V(\nu_c)$, $s(\nu_c)$, and $W(\nu_c)$ and estimate the effective bandwidth $\Delta \nu$ as:
\begin{equation}
    \Delta \nu = \frac{\int |W'(\nu) V(\nu)| d\nu}{s(\nu_c) V(\nu_c) W(\nu_c)}.
\end{equation}
These values are shown in \cref{tab:bandpass}, along with the nominal band center and bandwidth.
In this calculation, we used the $V$ spectrum at azimuth of $180^\circ$ and elevation of $45.5^\circ$.
Note that the effective band center of wafer 8.2.0 is higher than that of other wafers opposite to the trend of the nominal band center.
This could be explained by the shape of the bandpass of wafer 8.2.0 that shows a gradual edge.

\subsection{Circular polarization signal in the detector timestream}
\label{sec:detector signal model}

The expected detector signal is estimated by calculating the series of Mueller matrices of the optical elements of the receiver without neglecting the circular polarization component~\citep{Fujino2023}.
Making the dependence from the HWP angle explicit\footnote{The angle of the HWP is defined as the angle of the ordinary axis of the sapphire in the instrumental coordinates.}, $\theta=\omega t$, the detector timestream can be decomposed into three harmonics of the HWP rotation frequency:
\begin{equation}
d(t) = d_0(t) + d_2(t) + d_4(t)\;.
\end{equation}
The unmodulated component $d_0(t)$ mainly contains variations
of the intensity signal as:
\begin{equation}
d_0(t) = \Delta I\;.
\end{equation}
while $d_4(t)$ is the component modulated at 4 times the rotation frequency.
The sky linear polarization signal (Stokes $Q$ and $U$ defined in the instrument coordinates) appears in this component as:
\begin{equation}
d_4(t) = \mathrm{Re}[(Q+iU)\exp(4i\theta + 2i\phi)]\;.
\end{equation}
Here, we wrote explicitly the detector polarization angle $\phi$, which is defined as the detector angle in the instrument coordinate system.

Finally, $d_2(t)$ is the component modulated at 2 times the rotation frequency of the HWP.
It can be written as:
\begin{equation}
\begin{split}
d_2(t) = \mathrm{Re}[ & (\rho I+ i sV)\exp(2i\theta + 2i\phi) \\
& + \rho(Q + i U)\exp(2i\theta)]
\;.\end{split}\label{eq:d_2(t)}
\end{equation}
Here, $\rho$ is the band-averaged $\rho(\nu)$ of the HWP Mueller matrix.
The first term corresponds to conversion from the Stokes $I$ and $V$ to linear polarizations by the HWP.
The second term corresponds to signal converted from the linear polarization to unpolarized signal, and thus it has a $\theta$ dependence but not $\phi$.

To extract the signals of interest, we demodulate the detector signal using the
measured HWP angle $\tilde{\theta}$ and detector angle $\tilde{\phi}$.
We obtain demodulated timestreams for each harmonic component:
\begin{align}
\label{eq:d0f}
\tilde{d}_0(t) & = F_\mathrm{LPF}[d(t)] \approx \Delta I\;, \\
\tilde{d}_4(t) &= F_\mathrm{LPF}\large[ d(t) \cdot 2 \exp(-4i\tilde{\theta} - 2i\tilde{\phi})\large]\nonumber\\
\label{eq:d4f}
& \approx Q + iU\;,\\
\tilde{d}_2(t) &= F_\mathrm{LPF}\large[ d(t) \cdot 2 \exp(-2i\tilde{\theta} - 2i\tilde{\phi})\large]\nonumber\\
\label{eq:d2f}
& \approx \rho I + i sV + \rho(Q+iU)\exp(-2i\tilde{\phi})\;,
\end{align}
where $F_\mathrm{LPF}$ is a low-pass filter used to isolate a specific harmonic and exclude other harmonics.
The intensity and linear polarization signals are obtained from 
$\tilde{d}_0(t)$ and $\tilde{d}_4(t)$.
The circular polarization signal $sV$ is obtained from the imaginary part of $\tilde{d}_2(t)$. The additional $\rho(Q+iU)$ term can be subtracted using orthogonal detector pairs.

While this is the basic minimum-complexity model, we have to consider various types of instrumental non-ideality: the HWP synchronous signal, imperfections of other optical elements, variations of the detector time constant, and the effect of off-axis incidence 
(Appendix~\ref{sec:components}).

\section{Instrument and data}
\label{sec:data}

\subsection{Instrument}
\label{sec:instrument}
The \textsc{Polarbear} experiment observed the CMB from the Atacama Desert in Chile.
A cryogenic receiver \citep{arnold2012bolometric, kermish2012polarbear} is installed on the Huan Tran Telescope. The reflective telescope has a 2.5\,m-aperture off-axis Gregorian optics.
The HWP described in \cref{sec:HWP} is placed at the prime focus and is continuously rotated at 2.0\,Hz during observations \citep{Takakura2017JCAP}.
Before reaching the focal plane, light passes through a vacuum window, infrared filter, another cryogenic HWP, and three re-imaging lenses.

On the focal plane, there are seven detector arrays on silicon wafers cooled to 0.3\,K.
Each wafer comprises 182 transition-edge sensor bolometers with lenslet-coupled double-slot dipole antennas \citep{arnold2012bolometric}.
The observing frequency band is determined by bandpass filters which show small variations across the array.
In this analysis, we use bandpass measurements made at the site using a Fourier transform spectrometer \citep[FTS,][]{Matsuda2019}.
The properties are listed in \cref{tab:bandpass}.

\begin{deluxetable*}{lrrr|rrrrr}
\tablecaption{Detector bandpass per wafer. The effective band center $\nu_c$, effective bandwidth $\Delta \nu$, signal amplitude $V(\nu_c)$, leakage coefficient $s(\nu_c)$, and bandpass $W(\nu_c)$ estimated in \cref{sec:simulation} are shown.
We adopted the value of the $V$ spectrum at azimuth $=180^\circ$ and elevation$=45.5^\circ$ in this calculation.\label{tab:bandpass}}
\tablehead{
\colhead{Wafer} & \colhead{Band center} & \colhead{Bandwidth} & \colhead{Lower edge} & \colhead{$\nu_c$} & \colhead{$\Delta \nu$} & \colhead{$V(\nu_c)$} & \colhead{$s(\nu_c)$} & \colhead{$W(\nu_c)$} \\[-5pt]
 & \colhead{[GHz]} & \colhead{[GHz]} & \colhead{[GHz]} & \colhead{[GHz]} & \colhead{[GHz]} & \colhead{[mK]} & & 
}
\startdata
8.2.0 & 136.9 & 30.4 & 121.7 & 122.1 & 2.7 & 3.5 & 0.48 & 0.53 \\
10.1  & 142.1 & 31.8 & 126.2 & 121.1 & 3.0 & 7.1 & 0.49 & 0.21 \\
10.2  & 143.5 & 32.6 & 127.2 & 121.8 & 3.5 & 4.1 & 0.48 & 0.21 \\
10.4  & 144.0 & 32.2 & 127.9 & 121.7 & 4.1 & 4.6 & 0.48 & 0.14 \\
10.5  & 145.5 & 31.8 & 129.6 & 121.9 & 3.9 & 3.8 & 0.48 & 0.12 \\
9.4   & 146.9 & 32.8 & 130.5 & 122.1 & 3.7 & 3.4 & 0.48 & 0.13 \\
10.3  & 148.7 & 31.0 & 133.2 & 122.0 & 4.9 & 3.5 & 0.48 & 0.05 \\
\enddata
\tablecomments{Band center and bandwidth are from \citet{Matsuda2019}}
\end{deluxetable*}

\subsection{Observations}\label{sec:obs}
\textsc{Polarbear} observed the sky from January 2012 to January 2017.
In this analysis, we use 3 years of data covering a 670-square-degree patch obtained using the continuously rotating HWP, starting in July 2014.
This data set is the same used in the analysis presented in \citet[\citetalias{PB2022reanalysis} hereafter]{PB2022reanalysis}, encompassing the third, fourth, and fifth seasons listed in \cref{tab:dataset}.

As described in \citet[\citetalias{PB2020BB} hereafter]{PB2020BB}, the telescope scans can be classified by scan direction---rising, middle, and setting---to follow the sky rotation.
For each scan direction, we perform a set of right-going and left-going constant elevation scans at a scan speed of $0.4\degr/\mathrm{s}$.
The scan ranges are listed in \cref{tab:scanposition}.
The elevation of the middle scan is changed in steps each day.
The duration of one observation is approximately 1 hour.

\begin{deluxetable}{cccr}
\tablecaption{
Properties of the scan directions
\label{tab:scanposition}}
\tablehead{
& & & \colhead{Number of} \\[-8pt]
\colhead{Name} & \colhead{Azimuth range} & \colhead{Elevation} & \colhead{observations}
}
\startdata
\textit{Rising} \\
r & [$133\degr$, $156\degr$] & $30.0\degr$ & 1042 \\
\textit{Middle}\\
m0 & [$154\degr$, $206\degr$] & $45.5\degr$ &   93 \\
m1 & [$154\degr$, $206\degr$] & $47.7\degr$ &  107 \\
m2 & [$154\degr$, $206\degr$] & $50.0\degr$ &  104 \\
m3 & [$150\degr$, $209\degr$] & $52.2\degr$ &  148 \\
m4 & [$150\degr$, $209\degr$] & $54.4\degr$ &  132 \\
m5 & [$145\degr$, $214\degr$] & $56.6\degr$ &  118 \\
m6 & [$145\degr$, $214\degr$] & $58.9\degr$ &  116 \\
m7 & [$145\degr$, $214\degr$] & $61.1\degr$ &   94 \\
m8 & [$138\degr$, $222\degr$] & $63.3\degr$ &  148 \\
m9 & [$138\degr$, $222\degr$] & $65.5\degr$ &  120 \\
\textit{Setting} \\
s  & [$202\degr$, $227\degr$] & $35.2\degr$ & 1471
\enddata
\end{deluxetable}

\subsection{Data selection}\label{sec:data selection}
The data selection in this analysis is based on that of \citetalias{PB2022reanalysis}.
For data selection, we establish thresholds and eliminate bad data, including readings from detectors with no optical response or no calibration data, timestreams with high noise levels, scans containing glitches, and observations made with poor weather conditions.
In addition to the selection criteria of \citetalias{PB2022reanalysis}, we discard 9.5\% of observations because they are outliers in terms of slope values of the circular polarization signal estimated as described later.
We flag these observations where the deviation of the azimuthal slope from the average slope across all observations is larger than 10 times the median absolute deviation (MAD) for one or more wafers.

\begin{deluxetable}{lrrrrr}
\tablecaption{Number of 1-hour long observations\label{tab:dataset}}
\tablehead{
                 &                 &                   &                 & \colhead{PWV}      & \colhead{PWV} \\[-13pt]
\colhead{Season} & \colhead{Total} & \colhead{Daytime} & \colhead{Night} &                    &               \\[-13pt]
                 &                 &                   &                 & \colhead{$<$1\,mm} & \colhead{$\ge$1\,mm}}
\startdata
Third season & 943 & 196 & 747 & 564 & 379 \\
Fourth season & 1453 & 413 & 1040 & 875 & 578 \\
Fifth season & 1297 & 314 & 983 & 824 & 473 
\enddata
\end{deluxetable}

\section{Analysis} \label{sec:analysis}

The atmospheric circular polarization signal is expected to have cosine-like variation as a function of the azimuth angle (\cref{fig:sV sim}, top).
Since the azimuthal angle range of the scans (\cref{tab:scanposition}) is limited,
only part of the curve can be determined, hence we compute the derivative of the signal with respect to the azimuth angle, which tells us the slope of the curve at a given azimuth
(\cref{fig:sV sim}, bottom).
We determine the difference in the slope of the curve for different scan directions.
As shown in \cref{fig:sV sim}, we expect the slope for the middle direction to be small.
We thus use the slopes of the rising and setting directions for the quantitative evaluation and use the slope of the middle direction for demonstration and validation purposes.
Finally, we evaluate the dependence from the detector bandpass (\cref{tab:bandpass}) from simulations (\cref{sec:simulation}) to confirm the strong frequency dependent nature of the atmospheric circular polarization signal as the ability to shown in \cref{fig:model_freq_dep,fig:sV sim}.

The challenge in our analysis is the ability to control the contamination from ground-synchronous signals leaking from $I$, $Q$, and $U$ parameters.
As shown in \cref{fig:impact_leak} and discussed in Appendix~\ref{appendix:leakage}, we find that this systematic error can be larger than the atmospheric circular polarization signal.
However, through the polarization modulation described in \cref{sec:detector signal model}, we can independently obtain other Stokes parameters ($I$, $Q$, and $U$) and subtract the ground contamination from the circular polarization signal.
Here, we assume that the atmospheric circular polarization signal is constant over the 3 years covered by our data, whereas the ground-synchronous signal may vary owing to activities at the site, such as the construction of the Simons Array, which involves upgrading the experiment to include two telescopes to the north and south of the \textsc{Polarbear} telescope.
Moreover, we assume that the ground-synchronous signal is not circularly polarized.

\subsection{Calibration and preprocessing}\label{sec:caltod}
Calibrations are based on those of \citetalias{PB2020BB} and \citetalias{PB2022reanalysis}.
Here, we describe only where the calibration approach differs in the present analysis.
As shown in \cref{sec:detector signal model}, the circular polarization signal has a dependence from $(\theta + \phi)$, whereas for the polarization angle we calibrate the linear combination $(2 \theta + \phi)$.
Here, $\theta$ and $\phi$ represent the angle of HWP and the detector polarization angle, respectively.
We, therefore, calibrate the HWP angle $\theta$ independently 
from the linearly polarized astronomical source measurements with and without the HWP.
We also conduct an additional calibration of the polarization efficiency.
We consider the contributions of the secondary mirror and the cryogenic HWP and 
estimate the polarization efficiency to be 92\%.
We ignore the pointing offset of each detector and the beam effect because
the atmospheric circular polarization signal varies over a scale
larger than the focal plane field of view.

Each detector timestream is calibrated in Rayleigh--Jeans temperature units using a thermal source, whose effective temperature itself is calibrated using Jupiter observations.
Unlike \citetalias{PB2020BB} and \citetalias{PB2022reanalysis}, the calibrations made using the CMB power spectrum are not applied because the source spectrum is different from that of the CMB.
The effect of the detector time constant is corrected by deconvolving the time constant.
The calibrated timestream is demodulated and downsampled using the angle of the rotating HWP determined independently with an encoder.
We extract the second harmonic, $\tilde{d}_2(t)$, in addition to the zeroth and fourth harmonics, $\tilde{d}_0(t)$ and $\tilde{d}_4(t)$ as mentioned above.
The demodulated timestreams are filtered by second-order polynomials over the 1-hour observation to remove slow variations that are mainly caused by fluctuations of the focal plane temperature.
Note that each observation typically includes 70 right-going or left-going scans, thus the reduction of the atmospheric signal by this filtering scheme is negligible.
In contrast to \citetalias{PB2022reanalysis}, we don't apply the polynomial filter scan-by-scan in this analysis.

Next, we average detector timestreams among detectors that belong to the same wafer and decompose them.
Following \cref{eq:d2f}, the demodulated timestream of the second harmonic term of the $i$-th detector is modeled as:
\begin{equation}
\tilde{d}_{2,i}(t) \approx \rho I + i s V + \rho (Q + i U) \exp(-2i\tilde{\phi}_i)\;,
\end{equation}
where $\tilde{\phi}_i$ is the detector polarization angle.
We solve this equation by averaging multiple detectors with different polarization angles as:
\begin{widetext}
\begin{equation}
\begin{split}&
\left[\begin{matrix}
\rho I& 
s V&
\rho Q&
\rho U
\end{matrix}\right]^{\mathsf{T}} \approx \left[\begin{matrix}
\tilde{d}^{(2,2,\mathrm{Re})}(t) &
\tilde{d}^{(2,2,\mathrm{Im})}(t) &
\tilde{d}^{(2,0,\mathrm{Re})}(t) &
\tilde{d}^{(2,0,\mathrm{Im})}(t)
\end{matrix}\right]^{\mathsf{T}} \\
&= 
\left[\begin{matrix}
\sum_i w_i & 
0 &  
\sum_i w_i \cos 2\tilde{\phi}_i &  
\sum_i w_i \sin 2\tilde{\phi}_i \\  
0 & 
\sum_i w_i &  
\sum_i -w_i \sin 2\tilde{\phi}_i &  
\sum_i w_i \cos 2\tilde{\phi}_i \\  
\sum_i w_i \cos 2\tilde{\phi}_i &  
\sum_i -w_i \sin 2\tilde{\phi}_i &
\sum_i w_i & 
0 \\ 
\sum_i w_i \sin 2\tilde{\phi}_i &  
\sum_i w_i \cos 2\tilde{\phi}_i &
0 &
\sum_i w_i
\end{matrix}\right]^{-1} \cdot
\left[\begin{matrix}
\sum_i w_i \mathrm{Re}[\tilde{d}_{2,i}(t)] \\
\sum_i w_i \mathrm{Im}[\tilde{d}_{2,i}(t)] \\
\sum_i w_i \mathrm{Re}[\tilde{d}_{2,i}(t) \exp(2i\tilde{\phi}_i)] \\
\sum_i w_i \mathrm{Im}[\tilde{d}_{2,i}(t) \exp(2i\tilde{\phi}_i)]
\end{matrix}\right].\label{eq:decomposition}
\end{split}
\end{equation}
\end{widetext}
Here, the weights $w_i$ are calculated using the inverse variance of the detector white noise in order to privilege data with high S/N.
The detector angle $\phi_i$ can have four values separated by $\pi/4$
such that the weight matrix is nearly diagonal.
These timestreams are summed over the detector index $i$ within a wafer so that the timestreams $d^{(*)}(t)$ are wafer-averaged values.
The superscript of the decomposed timestreams in \cref{eq:decomposition}
shows
the harmonic term,
the detector polarization angle,
and whether it is the real or imaginary component (see, e.g., \cref{eq:d_2(t)}).
The circular polarization signal is extracted from the $(2,2,\mathrm{Im})$ component:
\begin{equation}
\tilde{d}^{(2,2,\mathrm{Im})}(t) \approx sV\;.
\end{equation}

\Cref{eq:decomposition} shows decomposition into four components. However, we could be considered more components with a different $\phi$-dependence.
These extra components can be caused by systematic effects, such as intensity-to-polarization leakage in the optics and detector nonlinearity.
We separate these systematic effects by extending the components for decomposition in the averaging method and reduce their contamination into the signal components used in the following analysis.
In practice, we decompose the zeroth, second, and fourth harmonics into three, eight, and eight components, respectively.
See Appendix \ref{sec:components} for more details.

The decomposed timestreams may contain systematic leakages from other modes.
We estimate the leakage and subtract it at a later stage. This step will be described in \cref{sec:leakage}.

\subsection{Estimation of the azimuthal slope} \label{sec:azfit}
The atmospheric circular polarization signal is expected to be sinusoidal as a function of azimuth. 
We fit each component of the wafer-average timestreams, $\tilde{d}^{(*)}(t)$, with a polynomial function of the telescope azimuth, $a(t)$, by minimizing:
\begin{equation}
\sum_t
w_\mathrm{scan} \Big[\tilde{d}^{(*)}(t) - \sum_{i=0}^{2} p_{i, \mathrm{obs}}^{(*)}\,\{\Delta a(t)\}^i\Big]^2\;,
\label{eq:azfit}
\end{equation}
where the per-subscan weight $w_\mathrm{scan} = \sum_i w_i$ is the total weight summed overall detectors, $\Delta a(t) = a(t) - a_0$ is the azimuth referenced to the scan center $a_0$, and $p_{i, \mathrm{obs}}^{(*)}$ denotes each of the polynomial coefficients. $p_{1, \mathrm{obs}}^{(2,2,\mathrm{Im})}$ corresponds to the azimuthal slope of the atmospheric circular polarization signal, $sV$.

We then take the average of the slopes from many observations for each scan direction, $d \in (\rm{r, m, s})$, and each wafer.
To downweight data with larger systematic leakage, we perform averaging in two steps.
First, we compute a weighted average of the slope value $p_{1,d,y}^{(*)}$ for each season, $y \in (3, 4, 5)$, using the inverse-variance weight $w_{\rm{obs}} = \sum_\mathrm{scan} w_\mathrm{scan}$.
We then compute the inverse-variance weighted average for all seasons, $p_{1,d}^{(*)}$,
using the statistical error (\cref{sec:error}) and the systematic error of the leakage subtraction (\cref{sec:leakagesystematics}).

We also estimate the azimuthal slope of the atmospheric circular polarization signal for each wafer and each scan direction from the simulations (\cref{sec:simulation}). We test their consistency by taking the ratio of the measured value and the simulation value.

\subsection{Leakage subtraction}\label{sec:leakage}
\begin{figure}
\centering
\includegraphics[width=\columnwidth]{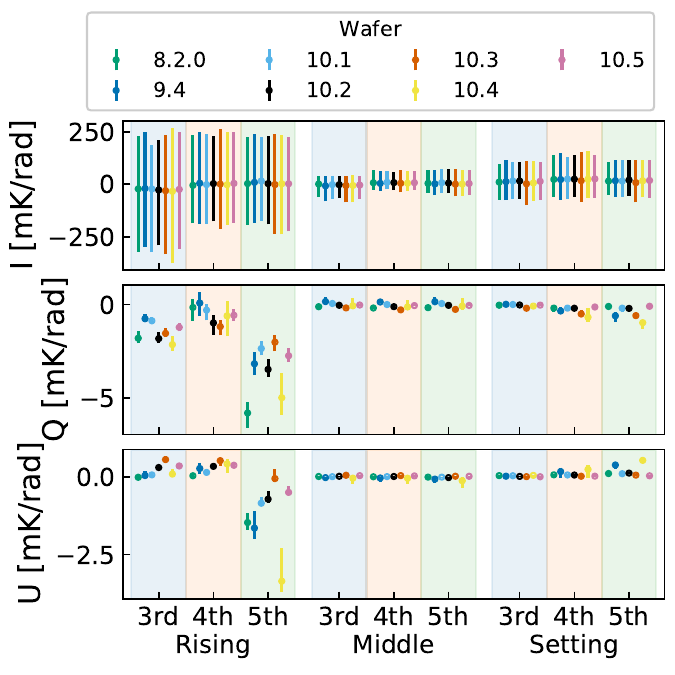}
\caption{Measured azimuthal slopes of the intensity and linear polarization signals for each wafer, season, and scan direction. The points show median values and the error bars show observation-by-observation variation.
The linear polarization signals shown are those after subtraction of the intensity leakage.
\label{fig:iqu variation}}
\end{figure}

Leakage of intensity and linear polarization into circular polarization is the most concerning source of systematic error.
We subtract the estimated leakage in our analysis.

We consider linear leakage from the intensity and linear polarization signals 
\footnote{This means that we consider only the primary cause of the leakage for each of $I$, $Q$, and $U$. To be precise, there could be more terms as discussed in Appendix~\ref{sec:components}. In addition, the $I$, $Q$, and $U$ include contributions from various sources, each of which may have a different leakage coefficient.}
expressed as:
\begin{equation}
\tilde{d}^{(2,2,\mathrm{Im})}(t) = sV + \lambda^{I\rightarrow sV} I + \lambda^{Q\rightarrow sV} Q + \lambda^{U\rightarrow sV} U\:,\label{eq:leakage_model_signal}
\end{equation}
where $\lambda^{X \to Y}$ is the leakage coefficient from each component $X$ to $Y$.
Models for the leakage coefficients based on the instrumental model are explained in Appendices~\ref{sec:components} and \ref{appendix:leakage}.

As intensity and linear polarization signals have large variations across seasons and non-zero values of the azimuthal slope as shown in \cref{fig:iqu variation},
their leakage may increase the uncertainty in the circular polarization measurements and bias the results.
Differences in the intensity signal could be due to atmospheric variations. Variations in the linear polarization signal
likely come from the far sidelobe pickup of ground equipment and human activities (see Appendix~\ref{sec:polleakagesubtraction}).

To minimize these systematic errors, we subtract the leakage using the estimated $I$, $Q$, and $U$ signals.
We first approximate \cref{eq:leakage_model_signal} by:
\begin{equation}
\begin{split}
\tilde{d}^{(2,2,\mathrm{Im})}(t) \approx sV
&+ \tilde{\lambda}^{I\rightarrow sV} \tilde{d}^{(0,0,\mathrm{Re})}\\
&+ \tilde{\lambda}^{Q\rightarrow sV} (\tilde{d}^{(4,2,\mathrm{Re})} - \tilde{\lambda}^{I\rightarrow Q} \tilde{d}^{(0,0,\mathrm{Re})}) \\
&+ \tilde{\lambda}^{U\rightarrow sV} (\tilde{d}^{(4,2,\mathrm{Im})}- \tilde{\lambda}^{I\rightarrow U} \tilde{d}^{(0,0,\mathrm{Re})})
\;,
\end{split}
\end{equation}
where $\tilde{\lambda}^{X\rightarrow Y}$ is the leakage coefficient estimated from observation data.
We then subtract the leakage by scaling the source signal $X$.
Here, we assume that variations in the atmospheric circular polarization signal for each scan direction are small (see \cref{sec:time_variation}) and that all variations correlated with the source signal are due to leakage.
We thus estimate the leakage coefficients such that the variance of the circular polarization signal after leakage subtraction is minimized.
In addition, we assume that this leakage subtraction removes the bias.
Here, we briefly describe the methods and results. Further details and discussions are presented in Appendix~\ref{appendix:leakage}.

We first estimate the coefficient of the leakage from the intensity signal, specifically the $(0,0,\mathrm{Re})$ component. 
We estimate the leakage coefficients for every 1-hour observation using the correlation of the wafer-averaged timestreams $\tilde{d}^{(*)}(t)$
and subtract the scaled template of the $(0,0,\mathrm{Re})$ component from the other components.
After this leakage subtraction, the variation in the slope of $p_{1,\mathrm{obs}}^{(2,2,\mathrm{Im})}$ decreases in all wafers except wafer 10.1.
In particular, the standard deviation decreases by approximately 90\% in wafer 10.2.
Moreover, the correlation between $p_{1,\mathrm{obs}}^{(0,0,\mathrm{Re})}$ and $p_{1,\mathrm{obs}}^{(2,2,\mathrm{Im})}$ decreases to $<7\%$ in all wafers.
In addition, the correlation between  $p_{1,\mathrm{obs}}^{(2,2,\mathrm{Im})}$ and the leakage coefficient of this intensity subtraction is $<10\%$.
Therefore, the systematic error due to the leakage of the intensity signal is well subtracted, and the residual has now a subdominant effect.

We then estimate the coefficients of linear polarization leakage, $(4,2,\mathrm{Re})$ and $(4,2,\mathrm{Im})$.
We estimate the leakage coefficients for each scan direction over all seasons to minimize the variance in the slope of the circular polarization signal $p_{1, \mathrm{obs}}^{(2,2,\mathrm{Im})}$ after subtracting scaled $p_{1, \mathrm{obs}}^{(4,2,\mathrm{Re})}$ and $p_{1, \mathrm{obs}}^{(4,2,\mathrm{Im})}$.
After this leakage subtraction, the standard deviation of the slope $p_{1,\mathrm{obs}}^{(2,2,\mathrm{Im})}$ decreases by approximately $10\%$ in wafer 8.2.0 and $80\%$ in wafer 10.4.

\subsection{Error estimation} \label{sec:error}
\begin{deluxetable*}{lccc}
\tablecaption{
List of uncertainties and correlations assumed
\label{tab:errorlist}}
\tablehead{
& \colhead{Correlated} & \colhead{Correlated} & \colhead{Correlated} \\[-8px]
& \colhead{among}      & \colhead{among}      & \colhead{among} \\[-8px]
\colhead{Type of uncertainty} & \colhead{directions} & \colhead{wafers} & \colhead{seasons}
}
\startdata
\textit{Measurement} \\
Statistical error & & & \\
Systematic error of linear polarization leakage subtraction & & & \\
Excess seasonal variation & & & \\
Systematic error of angle calibration & & & $\checkmark$ \\
Systematic error of absolute gain & $\checkmark$ & $\checkmark$ & $\checkmark$ \\
Systematic error of polarization efficiency & $\checkmark$ & & $\checkmark$ \\\hline
\textit{Simulation}\\
Systematic error of bandpass uncertainty & $\checkmark$ & & $\checkmark$ \\
Systematic error of the HWP & $\checkmark$ & $\checkmark$ & $\checkmark$ \\
Systematic error of temporal variation of the signal & $\checkmark$ & $\checkmark$ & $\checkmark$ \\
\enddata
\end{deluxetable*}
\vspace{-2\intextsep}
We estimate the uncertainties for both the measured and simulated values of the azimuthal slope of the atmospheric circular polarization signal for each wafer and scan direction. The uncertainties are summarized in \cref{tab:errorlist}.

For the measurement case, we consider the statistical uncertainty, the systematic uncertainty due to the leakage subtraction, the uncertainty in the polarization angle, the uncertainties in the absolute gain and polarization efficiency, and the extra-seasonal variations.
For the simulation case, we consider systematic uncertainties relating to the detector bandpass, the temporal variation in the atmospheric circular polarization, the uncertainty in the atmospheric model, and the uncertainty in the HWP model.

We estimate the measurement uncertainties in two steps as shown in \cref{sec:azfit}.
We first estimate the uncertainty for each season\footnote{For each wafer and scan direction}, including the statistical uncertainty, the systematic uncertainty in the leakage subtraction, and the extra-seasonal variation.
We then estimate the uncertainty for the average among seasons 
and add other systematic uncertainties in quadrature.

We consider that all simulation uncertainties are multiplicative. Therefore, we estimate fractional uncertainties relative to the simulation value for each source and take their quadratic sum.

Finally, the uncertainty in the ratio of the measurement to the simulation for each wafer and direction is estimated through the error propagation of both uncertainties.
The uncertainty in the ratio averaged over wafers and directions is also calculated through the error propagation considering the correlations as shown in \cref{tab:errorlist}.

\subsubsection{Statistical uncertainty}
We estimate the statistical uncertainty for each wafer and scan direction $d$ for each season $y$ using the sign-flip method.
We calculate:
\begin{equation}
p_{1,d,y,\mathrm{sf}}^{(2,2,\mathrm{Im})} 
= \frac{
  \sum_{\mathrm{obs}\in (d,y)} f_\mathrm{obs} w_\mathrm{obs} p_{1,\mathrm{obs}}^{(2,2,\mathrm{Im})}
}{
  \sum_{\mathrm{obs}\in (d,y)} w_\mathrm{obs}
}\;,
\end{equation}
where the set of random values $f_\mathrm{obs}=\{1$ or $-1\}$ is realized many times such that $\sum_{\mathrm{obs}\in (d,y)} w_\mathrm{obs} f_\mathrm{obs} \sim 0$. The standard deviation of the results is then taken.
Here, $p_{1,\mathrm{obs}}^{(2,2,\mathrm{Im})}$ is calculated after the leakage subtractions.

\subsubsection{Systematic uncertainty in the leakage subtraction}\label{sec:leakagesystematics}
Leakage subtraction with wrong estimations of the leakage coefficients, $\tilde{\lambda}^{I \to sV}$, $\tilde{\lambda}^{Q \to sV}$, and $\tilde{\lambda}^{U \to sV}$ may introduce systematic error.

For the leakage from intensity, originating from the large signal due to atmospheric fluctuation, we can accurately estimate the leakage coefficient for each observation, and the cause of its variation is detector nonlinearity (Appendix \ref{appendix:intensity leakage}).
We thus assume that this systematic error is negligible. 
This is supported by an analysis of correlation between $p_{1, \mathrm{obs}}^{(2, 2, \mathrm{Im})}$ and the intensity leakage coefficient in \cref{sec:leakage}, which is sensitive to this effect.

For leakage from linear polarization, the source of the leakage and its variation are not fully understood.
We estimate the statistical uncertainty of the measured leakage coefficient using the bootstrap method. 
In addition, we estimate the systematic uncertainty in the leakage coefficients from the seasonal variations of the estimated leakage (See \cref{fig:leakage}).
To account for the seasonal variation in the linear polarization signal shown in \cref{fig:iqu variation}, we evaluate the systematic error for each season and direction as:
\begin{equation}
\begin{split}
&\left[
\left(\tilde{\lambda}^{Q \to sV} \delta p_{1,d,y,\mathrm{stat}}^{(4,2,\mathrm{Re})}\right)^2
+\left(\tilde{\lambda}^{U \to sV} \delta p_{1,d,y,\mathrm{stat}}^{(4,2,\mathrm{Im})}\right)^2
\right.\\
&\left.
+\left(\delta \tilde{\lambda}^{Q \to sV}_\mathrm{stat} p_{1,d,y}^{(4,2,\mathrm{Re})}\right)^2
+\left(\delta \tilde{\lambda}^{U \to sV}_\mathrm{stat} p_{1,d,y}^{(4,2,\mathrm{Im})}\right)^2
\right.\\
&\left.
+\left(\delta \tilde{\lambda}^{Q \to sV}_\mathrm{sys} p_{1,d,y}^{(4,2,\mathrm{Re})}\right)^2
+\left(\delta \tilde{\lambda}^{U \to sV}_\mathrm{sys} p_{1,d,y}^{(4,2,\mathrm{Im})}\right)^2
\right]^{1/2}\;.
\end{split}
\end{equation}
Because of the seasonal variation in the linear polarization signals, we assume that this error is uncorrelated among wafers, directions, and seasons.
The final systematic uncertainty of the leakage subtraction for each wafer and scan direction
is estimated from the error propagation in the average among all seasons, which is approximately 10\% of the signal amplitude.

\subsubsection{Excess systematic variation between seasons}
We incorporate an additional systematic uncertainty for seasonal variations specific for each wafer and scan direction.
This uncertainty is adjusted so that the reduced chi-squared of the average among seasons becomes unity if the original residual chi-squared exceeds unity.
Here, we assume that this error is uncorrelated among seasons.
This procedure is applied to wafers 8.2.0 and 10.1, which are expected to see larger signals of atmospheric circular polarization and might be more sensitive to multiplicative systematic uncertainties.

\subsubsection{Polarization angle uncertainty}\label{sec:polangleerror}

In this work we follow the polarization angle calibration of \citetalias{PB2020BB}, with an associated uncertainty of $0.22^\circ$, which is negligibly small for this analysis.
This calibration determines the linear combination of $2\tilde{\theta} + \tilde{\phi}$ whereas, as described in \cref{sec:caltod}, the circular polarization signal depends on $\theta+\phi$.
We thus need to estimate the HWP angle $\tilde{\theta}$ independently from the angle calibration.
The absolute value, or the calibrated zero point of $\tilde{\theta}$, contains a systematic error.
This angular error affects the phase of the demodulated signal and cause intensity leakage $\rho I$ into the circular polarization signal $sV$, as described by:
\begin{equation}
\begin{split}
\tilde{d}^{(2,2,\mathrm{Re})} & \approx \rho I \cos 2\Delta - s V \sin 2 \Delta\;, \\
\tilde{d}^{(2,2,\mathrm{Im})} & \approx sV \cos 2\Delta + \rho I \sin 2 \Delta\;, 
\end{split}
\end{equation}
where $\Delta=\tilde{\theta} - \theta$ is the absolute angular error of the HWP angle.

We evaluate $\Delta$ using the polarized 
second harmonic of the HWP synchronous signal (2f HWPSS):
$p_{0, \mathrm{obs}}^{(2,2,\mathrm{Re})}$ and $p_{0, \mathrm{obs}}^{(2,2,\mathrm{Im})}$ in \cref{eq:azfit}.
As described in the optical model of the HWP in \cref{sec:HWP}, 
the main sources of polarized 2f HWPSS are the differential transmission, reflection, and emission of the HWP, as well as the circular polarization signal, $s V$.
In an ideal case, the circular polarization signal appears in the imaginary part of the polarized 2f HWPSS ($p_{0, \mathrm{obs}}^{(2,2,\mathrm{Im})}$) and differential transmission, reflection, and emission of the HWP appear in the real part ($p_{0, \mathrm{obs}}^{(2,2,\mathrm{Re})}$) but the angular error changes the phase and cause a mixing.
Among these, we expect that the variation in the differential transmission ($\rho I$) is the largest component owing to the variation in the observation weather.
We thus can estimate the potential angular error
$\Delta$ by performing the principal component analysis on the $2 \Delta$-rotated slopes of the 2f HWPSS to determine the phase that maximizes the variation in the $\rho I$ term.

We estimate this phase rotation of the polarized 2f signal for each wafer, scan direction, and season.
We find that there are systematic variations in the phase rotation of the polarized 2f signal ($2\Delta$) between $0\degr$ and $8\degr$.
Therefore, we estimate the systematic 2f-phase uncertainty $2\delta\Delta = 4\degr$ and the systematic uncertainty of the circular polarization signal as $2|\tilde{p}^{(2,2,\mathrm{Re})}_{1,d}|\delta\Delta$.
To be conservative, we take the maximum value among directions for each wafer and treat this uncertainty as being uncorrelated among directions and wafers.
The corresponding systematic uncertainties in the slope of the $s V$ signal are between 10\% and 60\% of the signal.

\subsubsection{Absolute gain and polarization efficiency}
Here, we consider multiplicative uncertainties in the measurements.
We use the same absolute gain value used in the linear polarization analysis \citepalias{PB2020BB}, which is calibrated with planet observations.
The fractional uncertainty of the absolute gain calibration is approximately 2\%.
We assume this error to be correlated among all wafers and directions.

The polarization efficiency also has uncertainty.
We estimate this uncertainty from the variation due to the detector position and the difference between wafers due to the bandpass difference.
The fractional uncertainty is approximately 5\%.
We assume this error to be uncorrelated between wafers.

\subsubsection{Bandpass uncertainty}
Our bandpass measurements are also affected by uncertainty \citep{Matsuda2019} causing a systematic error in the band average (see Equation \ref{eq:band averaged sV}).
We evaluate this by looking at variations in the detector response within the wafer as follows.
First, we generate 5000 random realizations of the bandpass for each wafer using the bootstrap method.
We then calculate the slope of the band-averaged $sV$ signal following the same approach as in \cref{sec:simulation} using a bandpass that includes this noise. We then compute
the standard deviation among realizations to obtain the uncertainty.
The fractional uncertainty is at the level of a few percent.

We also worry that the bandpass measurement contains a systematic error in frequency scaling.
To evaluate this, we calculate the slope of the $sV$ signal by re-scaling the frequency of the bandpass and compare it with the value obtained without re-scaling.
We define the systematic error as the maximum slope difference we obtain when re-scaling the frequency within the $1\,\mathrm{GHz}$ resolution of the FTS measurement.
These fractional uncertainties are typically 8\%, 14\% in the worst case.

We assume these errors to be correlated among directions but not among wafers because they depend on the shape of the bandpass and possibly on the detector position on the focal plane.

\subsubsection{HWP model uncertainty}
We previously estimated in \citet{Fujino2023} that the uncertainty in the HWP leakage parameter $s$ is less than 3\%.

\subsubsection{Temporal variation in the atmospheric circular polarization signal}\label{sec:time_variation}
The direction and amplitude of Earth's magnetic field change year by year.
In the simulation, we calculate the expected atmospheric circular polarization using the direction and amplitude of Earth's magnetic field in the middle of the observation period using the enhanced magnetic model 2015 \citep[EMM2015,][]{EMM2015}.
To estimate the effects of the temporal variation of Earth's magnetic field on the sV signal, we calculate the direction and amplitude of Earth's magnetic field at the beginning, middle, and end of the observation period.
The fractional difference in amplitude is 1\%, and the difference in the direction is approximately $0.5^\circ$ in this period.
The impact of this temporal variation on the slope of the $sV$ signal is approximately 5\% for the middle direction, less than 2\% for the rising direction, and less than 1\% for the setting direction.
This dependence on the observation direction comes from the variation in the direction of Earth's magnetic field.
The effect on the middle direction is relatively pronounced owing to the small absolute value.
However, we use the data in the middle direction as a reference only and do not use them in the final evaluation, rendering this a minor effect.

Temperature and pressure are also time-varying parameters.
The standard deviation of the pressure variations observed at the \textsc{Polarbear} site is less than 1\%, and the effect on the signal is 1\% or less.
The seasonal temperature variation is typically $\sim 30\,\rm{K}$ at the observation site.
According to Table 5 in \cite{Spinelli2011}, the effect of temperature variation on Zeeman emission is less than $3\,\rm{\mu K}$, which corresponds to 5\% of the signal.
Furthermore, we assume this does not create azimuthal variation.
Therefore, we can confidently disregard this systematic effect.

A side effect of this variation of the atmospheric circular polarization signal may appear in the leakage subtraction (\cref{sec:leakage}), where we estimate the leakage coefficients by minimizing the variance. However, since the estimated temporal variation is smaller than our statistical variation, the bias should be smaller than the current estimate of the systematic uncertainty due to the leakage subtraction (\cref{sec:leakagesystematics}).

\subsubsection{Atmospheric emission model uncertainty}

In \citetalias{class2020}, they found discrepancies of $\sim 20\%$ in signal amplitude between data and simulation.
It was concluded that this was due to simulation inaccuracy of the atmospheric spectrum frequencies away from the oxygen resonance line.
In our case, as we observed the signal close to the resonance line, this effect is considered to be less than 20\%.

\subsection{Validation}

We validate the estimated azimuthal slope of the circular polarization component by comparing the results obtained from subsets of data.
We consider six subsets for the
right-going scan, 
left-going scan,
high precipitable water vapor
($\mathrm{PWV}>1\,\mathrm{mm}$), 
low PWV ($\mathrm{PWV}\leq1\,\mathrm{mm}$),
daytime, 
and night.
In the scan direction splits, we test systematic errors due to variations in the focal plane temperature.
In the PWV splits, we can examine systematic errors arising from intensity leakage or variations in the time constant.
In the daytime and night splits, we can investigate systematic errors related to ground temperature variations or far sidelobe pick-up of the Sun.

In the scan direction splits, we subtract the intensity leakage and estimate the azimuthal slope for each observation using the corresponding scans. In the other splits, we use the same coefficients of intensity leakage as used in the main analysis for each subset of observations.
We use the same coefficients of linear polarization leakage as used in the main analysis for all cases.
We use the measurement uncertainty estimated from all data including the statistical uncertainty and the systematic uncertainties from leakage subtraction, excess seasonal variations, and polarization angle uncertainty.
As shown in \cref{tab:sys_errs}, systematic uncertainties are dominant in our observation compared to statistical uncertainty.

\begin{figure*}
\centering
\includegraphics[width=\textwidth]{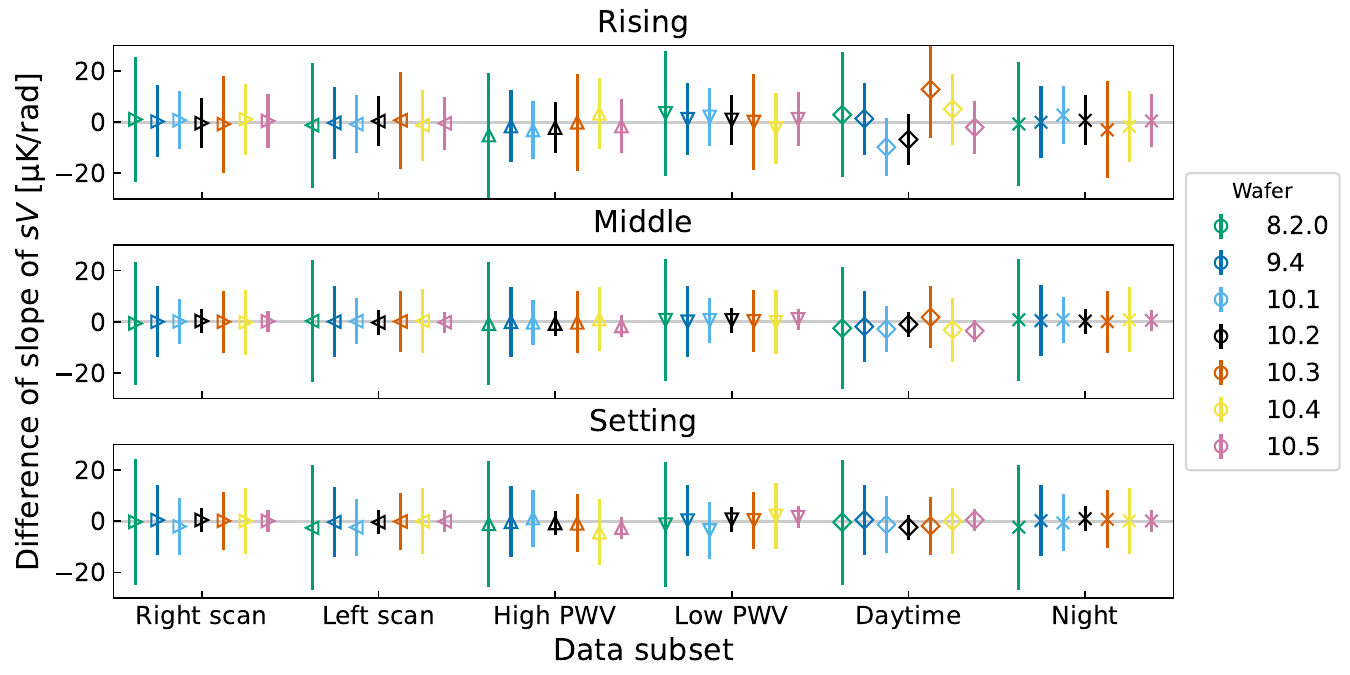}
\caption{Deviation of measurements obtained from each subset of data relative to the full data analysis. Panels from top to bottom show values for rising, middle, and setting scan directions. The color of the points denotes the wafer and the symbol of the points indicates the subset of data.\label{fig:validation}}
\end{figure*}

\Cref{fig:validation} shows the results.
All data fall within $0.9 \sigma$ of zero.
The variation in each subset is affected by systematic error in addition to statistical error.
Since these discrepancies are smaller than the uncertainties estimated above, we conclude that the slopes from all subsets align with the full data analysis, and there are no other appreciable systematic uncertainties.

\section{Results}\label{sec:results}

\begin{figure*}
    \centering
    \includegraphics[width=\textwidth]{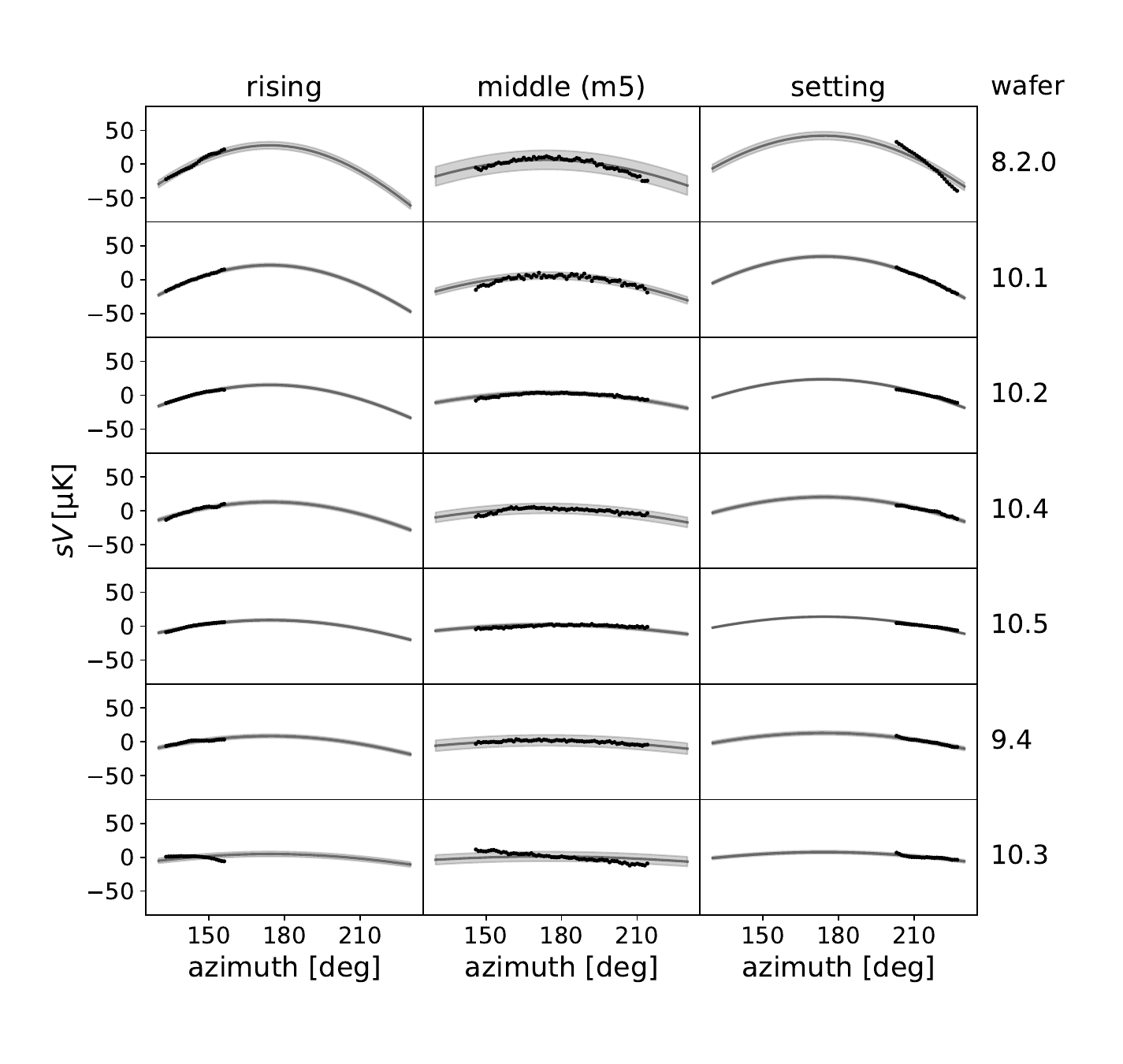}
    \caption{Binned azimuthal profiles for all wafers and scan directions (black points).
    Average values have been subtracted.
    Simulated $sV$ values are overlaid (gray lines).
    The gray bands show roughly estimated one-sigma regions obtained as half of the root sum square of errors in \Cref{tab:sV_results} times the azimuthal range.
    Note that there is an outlier for the setting direction, but the weight of this point is low and the effect on the calculation of the azimuthal slope is negligible. Therefore, we do not show this point.}
    \label{fig:sV_1d_map}
\end{figure*}
\Cref{fig:sV_1d_map} shows binned azimuthal profiles plotted together with the simulated values.
The data points in the rising and setting directions have a positive and negative azimuthal dependence as expected.
Moreover, we find that data points in the middle direction peak around an azimuth of 180 degrees.
The data points have an azimuthal dependence compatible with that of the simulation. 
However, some wafers and directions appear to be slightly different from the simulation.
The row of panels is ordered by the lower edge of the bandpass, and we see that the higher the lower edge of the bandpass is, the smaller the amplitude of the sinusoidal curve is found to be.
Furthermore, we show roughly estimated errors as the shaded regions in \cref{fig:sV_1d_map}.
These errors are calculated as the azimuthal range multiplied by half of the root sum square of all errors estimated in \cref{sec:error}.
We can see that almost all of the data points in the rising and middle directions are within these error bands.
\begin{figure}
    \centering
    \includegraphics[width=\columnwidth]{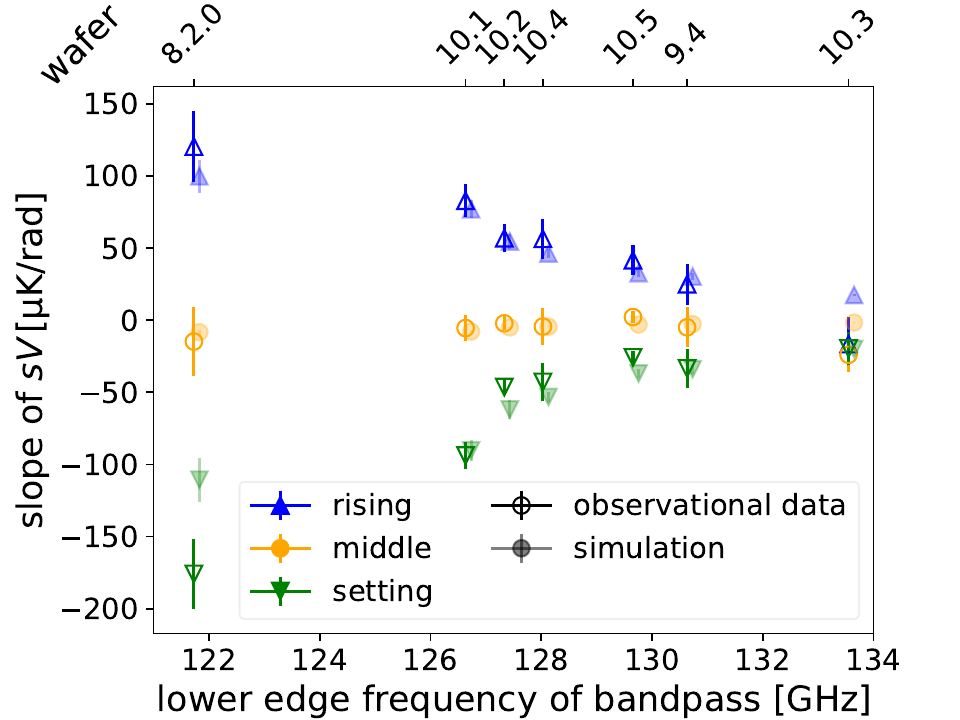}
    \caption{
Slope of the apparent circular polarization signal versus the lower passband frequency of each wafer.
Both observational and simulation data are shown.
    }
    \label{fig:sV_results1}
\end{figure}

\begin{table*}[]
    \centering
    \caption{Slope values, observational data/simulation data ratios, and statistical error calculated in \Cref{sec:analysis}.
    All slope values are in units of $\rm{\mu K /rad}$.}
    {
    \tabcolsep = 3pt
    \begin{tabular}{lc|ccccccc}
        \hline
        \hline
        & & \multicolumn{7}{c}{Wafer} \\
        & & 8.2.0 & 9.4 & 10.1 & 10.2 & 10.3 & 10.4 & 10.5 \\
        \hline
        Azimuthal slope
        & rising & $120.4$ & $25.0$ & $82.9$ & $56.9$ & $-16.5$ & $56.4$ & $41.6$ \\
        of observational data
        & setting & $-175.9$ & $-33.5$ & $-93.9$ & $-46.4$ & $-19.6$ & $-42.8$ & $-25.8$ \\
        \hline
        Azimuthal slope
        & rising & $99.7$ & $30.3$ & $76.9$ & $54.6$ & $17.5$ & $46.3$ & $32.6$ \\
        of simulation
        & setting & $-110.5$ & $-34.0$ & $-90.1$ & $-61.9$ & $-20.0$ & $-53.2$ & $-37.0$ \\
        \hline
        Observational data/simulation & rising & $1.21 \pm 0.02$ & $0.82 \pm 0.09$ & $1.08 \pm 0.03$ & $1.04 \pm 0.08$ & $-0.95 \pm 0.17$ & $1.22 \pm 0.12$ & $1.28 \pm 0.13$ \\
        data and statistical error & setting & $1.59 \pm 0.01$ & $0.99 \pm 0.02$ & $1.04 \pm 0.01$ & $0.75 \pm 0.01$ & $0.98 \pm 0.06$ & $0.81 \pm 0.02$ & $0.70 \pm 0.02$ \\
        \hline
    \end{tabular}
    }
    \label{tab:sV_results}
\end{table*}

\begin{table}[]
    \centering
    \caption{List of errors in the averaged ratio.}
    \begin{tabular}{lr}
        \hline
        \hline
        Type of uncertainty & Value \\\hline
        Statistical error & $0.01$ \\
        \hline
            Systematic error in &  \multirow{2}{*}{$0.01$}\\
            \hspace{1em}linear polarization leakage subtraction & \\
        Excess seasonal variation & $0.02$ \\
        Systematic error in angle calibration & $0.05$ \\
        Systematic error in absolute gain & $0.02$ \\
        Systematic error in polarization efficiency & $0.02$ \\
        Systematic error in bandpass uncertainty & $0.04$ \\
        Systematic error related to the HWP & $0.03$ \\
        Systematic error in temporal variation of the signal & $< 0.01$ \\
        Atmospheric emission model uncertainty & $\ll 0.20$ \\
        \hline
    \end{tabular}
    \label{tab:sys_errs}
\end{table}

We summarize the results of the azimuthal slope of the circular polarization signal $sV$ for each wafer and scan direction in \Cref{fig:sV_results1} and \cref{tab:sV_results}.
We find a signal in both rising and setting observations, with opposite signs as expected.
The amplitude of the signal is typically 20--100$\,\mathrm{\mu K / rad}$ depending on the lower edge of the passband frequencies.
Moreover, \Cref{fig:sV_results1} shows the azimuthal slope derived from the simulation.
The estimated slopes are largely consistent with the simulation, particularly in terms of how their signs depend on azimuth and how their amplitudes relate to the wafers' bandpass.
However, there are discrepancies, which are discussed below.

\begin{figure}
    \centering
    \includegraphics[width=\columnwidth]{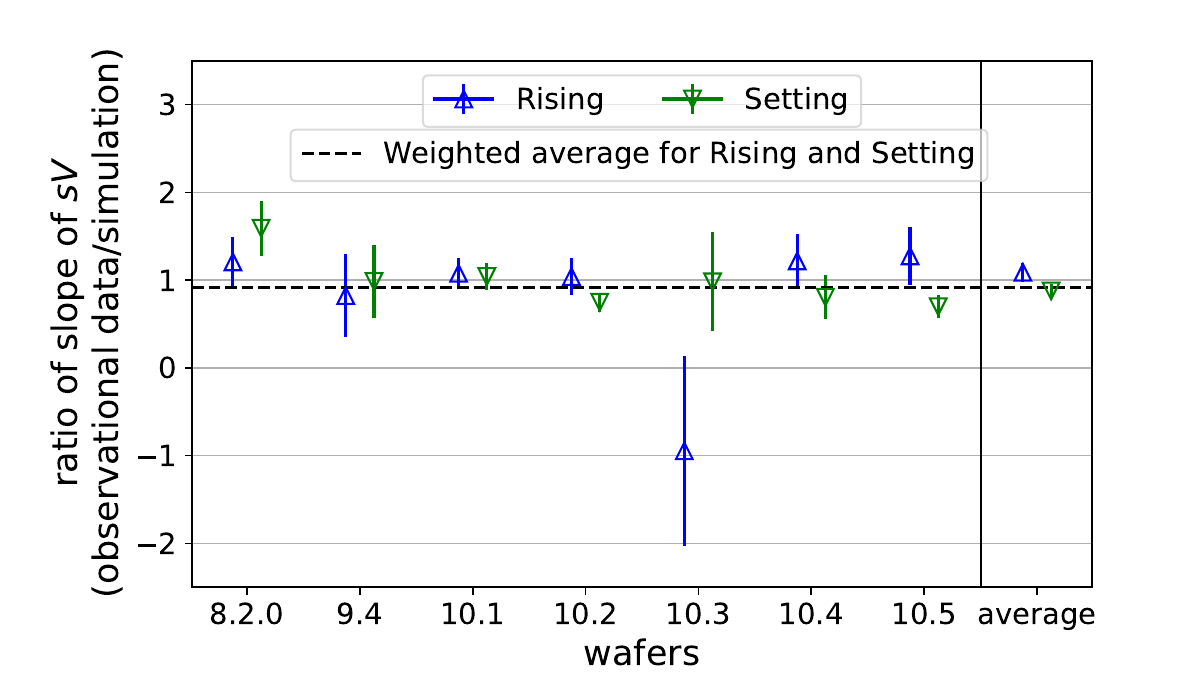}
    \caption{Ratio of the slope of $sV$ obtained from observational data to that obtained in simulation for each wafer in rising and setting directions.
    Also shown are the individual weighted-average values for each scan direction (at the right end), and the overall weighted-average value (the black dashed line).
    }
    \label{fig:sV_compare}
\end{figure}
In \cref{fig:sV_compare}, we show the ratio of the slope values between data and simulation for each wafer in each of the rising and setting directions.
We also show the weighted average over all wafers for each direction.
The chi-squared values for the weighted average of these ratios for the rising and setting directions are 6.7 and 7.0, respectively, with six degrees of freedom.
These chi-squared values correspond to p-values of 35\% and 32\%, respectively.
The two average values of the data-to-simulation ratio of the slope are $1.09 \pm 0.02 \mathrm{(stat)} \pm 0.11 \mathrm{(sys)}$ and $0.88  \pm 0.01 \mathrm{(stat)} \pm 0.08 \mathrm{(sys)}$ for rising and setting directions, respectively.

We then average the ratio for the rising and setting directions.
We show the statistical and systematic errors of this average in \cref{tab:sys_errs}.
The chi-squared is 14.3 with 13 degrees of freedom, and the p-value is 35\%.
The average is $0.92 \pm 0.01 \mathrm{(stat)} \pm 0.07 \mathrm{(sys)}$.
This ratio corresponds to a brightness temperature of $3.8\,\mathrm{m K}$ for a representative detector on wafer 10.2, using the values shown in \cref{tab:bandpass}.

\section{Conclusion} \label{sec:conclusion}
In this paper, we reported the observation of atmospheric circular polarization at a new frequency band of $150\,\rm{GHz}$ (effective frequency $121.8\,\mathrm{GHz}$) using \textsc{Polarbear} data obtained using a continuously rotating HWP.
The sign of the detected signal changes depending on the directions (rising or setting) as expected.
We found the largest slope of $120.4 \pm 2.4 \mathrm{(stat)} \pm 25.1 \mathrm{(sys)}\, \rm{\mu K/rad}$ (rising) and $-175.8 \pm 1.4 \mathrm{(stat)} \pm 25.8 \mathrm{(sys)}\, \rm{\mu K/rad}$ (setting) for the detectors on wafer 8.2.0, which shows the lowest lower edge of the frequency bandpass across wafers.
We also found a dependence on small shifts of the detector bandpass.
We compared these values with results from simulations.
We found consistency in the bandpass and scan direction dependence between the observational data and the simulations with a p-value of 35\%.
We took the ratio of the slopes between data and simulation.
The average ratio, for both rising and setting directions, was $0.92 \pm 0.01 \mathrm{(stat)} \pm 0.07 \mathrm{(sys)}$.
CLASS reported that the ratio at $40\,\mathrm{GHz}$ is approximately 80\% \citep{class2020}.
Our result favors the atmospheric origin of the signal.
This could be because we observed the signal close to the resonance line of the oxygen molecule, and the uncertainty of the model they considered did not affect our result.
In future investigations, we will need to better calibrate the absolute angle of the HWP and the absolute frequency of the bandpass measurement.

This work is the first demonstration of the measurement of circular polarization exploiting the continuously rotating HWP imperfections, which cause a conversion of the circular polarization signal into the imaginary part of the second harmonic signal at the HWP rotation frequency.
Although this study focused on measuring atmospheric circular polarization, 
our method is applicable to cosmological circular polarization.
Polarization modulation using a rotating HWP is currently being widely adopted and will continue to be used in future CMB experiments, including the Simons Array and Simons Observatory experiments \citep{hwp_chara_PB2,hwp_chara_SO}.
LiteBIRD, a satellite experiment searching for CMB $B$-mode polarization, also plans to install a continuously rotating HWP on each of its telescopes \citep{litebird2023}.
In these experiments, our method makes it possible to probe the circular polarization signal using data from linear polarization measurement.
The expected noise levels of these experiments for circular polarization are better than those of current experiments \citep[e.g., ][]{class_clvv2024}.

We discussed the systematic uncertainties affecting circular polarization observations and found that the leakage from the intensity and linear polarization components, the uncertainty of the polarization angle, and the uncertainty of the detector bandpass have a large effect on the results.
These systematic effects should be correctly evaluated in future cosmological circular polarization observations.
Our methodology and discussion in \cref{sec:analysis} will be helpful in this regard.

This work is supported by
the Gordon and Betty Moore Foundation Grant No. GBMF7939,
the Simons Foundation Grant No. 034079,
and the World Premier International Research Center Initiative (WPI) of MEXT, Japan.
In Japan, this work was supported by JSPS KAKENHI grant Nos. 18H05539 and 19H00674, and the JSPS Core-to-Core Program JPJSCCA20200003.
Work at LBNL is supported in part by the U.S. Department of Energy, Office of Science, Office of High Energy Physics, under Contract No. DE-AC02-05CH11231.
Calculations were performed on the Central Computing System, owned and operated by the Computing Research Center at KEK,
and the National Energy Research Scientiﬁc Computing Center, which is supported by the Department of Energy under Contract No. DE-AC02-05CH11231.
C.B. acknowledges partial support from the Italian Space Agency LiteBIRD Project (ASI Grants No. 2020-9-HH.0 and 2016-24-H.1-2018), the InDark and LiteBIRD Initiative of the National Institute for Nuclear Physics, and the RadioForegroundsPlus Project HORIZON-CL4-2023-SPACE-01, GA 101135036.
Y.C. acknowledges support from JSPS KAKENHI Grant Number 24K00667.
M.H. acknowledges support from JSPS KAKENHI Grant Number 22H04945.
C.R. acknowledges support from the Australian Research Council's Discovery Projects scheme (DP210102386).
S.T. acknowledges support from MEXT KAKENHI (JP18H04362) and JSPS KAKENHI (JP20K14481, JP18J02133, JP14J01662).
We thank Glenn Pennycook, MSc, from Edanz (https://jp.edanz.com/ac) for editing a draft of this manuscript.
We thank Tommaso Ghigna for improving the draft of this manuscript.

\appendix
\section{Detector timestreams with full instrumental nonidealities}\label{sec:components}
In \cref{sec:detector signal model}, we described the detector signal with ideal instruments except for nonidealities of the HWP. Here, we explain the model with nonidealities of other optical components.
The Stokes vector after the HWP is:
\begin{equation}
\left[\begin{matrix}I'\\Q'\\U'\end{matrix}\right] = \left[\begin{matrix}
I + \rho Q \cos 2\theta - \rho U \sin 2\theta\\
Q\cos 4\theta - U \sin 4\theta + \rho I \cos 2\theta - sV \sin 2\theta \\
-Q\sin 4\theta - U \cos 4\theta - \rho I \sin 2\theta - sV \cos 2\theta
\end{matrix}\right]\;.
\end{equation}

We first consider the optical elements on the sky side of the HWP.
The primary mirror is the sole instrument on this side in the \textsc{Polarbear} optics.
It slightly polarizes the incident unpolarized signal and also emits a polarized signal, expressed as:
\begin{equation}
Q \rightarrow Q + \lambda_1 I + Q_1\;.
\end{equation}

We next consider the polarized emission and reflection from the HWP. For our single-layer HWP, the polarization angle should be aligned with the birefringence axis. These effects can be represented by:
\begin{equation}
\rho I \rightarrow \rho I + Q_\mathrm{h}\;.
\end{equation}

We next consider the optical elements on the detector side of the HWP, including the secondary mirror, infrared filters, a cryogenic HWP, and three re-imaging lenses.
Note that the cryogenic HWP is not rotated during the observation period in this analysis.
The secondary mirror has an effect similar to that of the primary mirror, and we thus have:
\begin{equation}
Q' \rightarrow Q' + \lambda_2 I' + Q_2\;.
\end{equation}
The contribution of the other optics is modeled using a Mueller matrix as:
\begin{equation}
\left[\begin{matrix}I''\\Q''\\U''\end{matrix}\right] = \left[\begin{matrix}
1 & \gamma_Q & \gamma_U \\
\lambda_Q & \epsilon + \xi_+ & \xi_\times \\
\lambda_U & \xi_\times & \epsilon - \xi_+
\end{matrix}\right]
\left[\begin{matrix}I'\\Q'\\U'\end{matrix}\right]\;.
\end{equation}
Here, $\lambda_Q$ and $\lambda_U$ are the intensity-to-polarization ($I\rightarrow Q/U$) leakage, $\gamma_Q$ and $\gamma_U$ are the polarization-to-intensity ($Q/U\rightarrow I$) leakage, $\epsilon$ is the polarization efficiency, and $\xi_+$ and $\xi_\times$ represent the polarization asymmetry.
Here, $\epsilon \sim 1$ and other nonideality parameters should be at the level of a few percent.

Finally, a detector measures the signal along its polarization direction as:
\begin{equation}
d = I'' + Q'' \cos 2\phi + U'' \sin 2\phi\;.
\end{equation}
We decompose the result into 15 terms according to the modulation by $\theta$ and $\phi$ as:
\begin{equation}
\begin{split}
d = & d^{(0,0,\mathrm{Re})} \\
& + d^{(0,2,\mathrm{Re})} \cos 2\phi - d^{(0,2,\mathrm{Im})} \sin 2\phi\\
& + d^{(2,0,\mathrm{Re})} \cos 2\theta - d^{(2,0,\mathrm{Im})} \sin 2\theta\\
& + d^{(2,2,\mathrm{Re})} \cos (2\theta + 2\phi) - d^{(2,2,\mathrm{Im})} \sin (2\theta + 2\phi)\\
& + d^{(2,-2,\mathrm{Re})} \cos (2\theta - 2\phi) - d^{(2,-2,\mathrm{Im})} \sin (2\theta - 2\phi)\\
& + d^{(4,0,\mathrm{Re})} \cos 4\theta - d^{(4,0,\mathrm{Im})} \sin 4\theta\\
& + d^{(4,2,\mathrm{Re})} \cos (4\theta + 2\phi) - d^{(4,2,\mathrm{Im})} \sin (4\theta + 2\phi)\\
& + d^{(4,-2,\mathrm{Re})} \cos (4\theta - 2\phi) - d^{(4,-2,\mathrm{Im})} \sin (4\theta - 2\phi),
\end{split}
\end{equation}
where $d^{(a,b,\mathrm{Re})}$ and $d^{(a,b,\mathrm{Im})}$ are components with the $\cos(a\theta + b\phi)$ term and $\sin(a\theta + b\phi)$ term, respectively. The components are calculated as follows.
The unmodulated unpolarized component including the intensity signal is calculated according to:
\begin{equation}
d^{(0,0,\mathrm{Re})} = I + \gamma_Q (Q_2 + \lambda_2 I)\;.
\end{equation}
The unmodulated polarized components are calculated according to:
\begin{align}
d^{(0,2,\mathrm{Re})} &= \epsilon (Q_2 + \lambda_2 I) + \lambda_Q I + \xi_+ (Q_2 + \lambda_2 I)\;,\\
d^{(0,2,\mathrm{Im})} &= -\lambda_U I - \xi_\times (Q_2 + \lambda_2 I)\;.
\end{align}
The main contribution should be the polarized signal from the secondary mirror.
The 2f-modulated unpolarized components are calculated according to:
\begin{align}
d^{(2,0,\mathrm{Re})} &= \rho (Q + Q_1 + \lambda_1 I) + \gamma_Q (\rho I + Q_\mathrm{h}) - \gamma_U s V\;, \\ 
d^{(2,0,\mathrm{Im})} &= \rho U + \gamma_U (\rho I + Q_\mathrm{h}) + \gamma_Q s V\;.
\end{align}
The dominant term should be $\rho Q_1$, the polarized emission from the primary mirror converted to intensity by the HWP.
The 2f-modulated polarized components including the circular polarization signal are calculated according to:
\begin{align}
d^{(2,2,\mathrm{Re})} =\:& \epsilon (\rho I + Q_\mathrm{h}) \nonumber\\
&+ \frac{1}{2}(\lambda_Q + (\epsilon + \xi_+)\lambda_2)\rho(Q + Q_1 + \lambda_1 I)\nonumber\\
&+ \frac{1}{2}(\lambda_U + \xi_\times \lambda_2)\rho U\;,\\
d^{(2,2,\mathrm{Im})} =\:& \epsilon s V \nonumber\\
&- \frac{1}{2}(\lambda_U + \xi_\times\lambda_2)\rho(Q + Q_1 + \lambda_1 I)\nonumber\\
&+ \frac{1}{2}(\lambda_Q + (\epsilon + \xi_+)\lambda_2)\rho U\;.
\end{align}
The dominant term is $\epsilon \rho I$, the leakage from intensity to polarization at the HWP.
The 2f-modulated polarized components, but with a rotation direction opposite that for $d^{(2,2,\mathrm{Re})}$ and $d^{(2,2,\mathrm{Im})}$, are calculated according to:
\begin{align}
d^{(2,-2,\mathrm{Re})} =\:& \xi_+ (\rho I + Q_\mathrm{h}) - \xi_\times s V \nonumber\\
&+ \frac{1}{2}(\lambda_Q + (\epsilon + \xi_+)\lambda_2)\rho(Q + Q_1 + \lambda_1 I)\nonumber\\
&- \frac{1}{2}(\lambda_U + \xi_\times \lambda_2)\rho U\;,\\
d^{(2,-2,\mathrm{Im})} =\:& \xi_\times (\rho I + Q_\mathrm{h}) + \xi_+ s V \nonumber\\
&+ \frac{1}{2}(\lambda_U + \xi_\times\lambda_2)\rho(Q + Q_1 + \lambda_1 I)\nonumber\\
&+ \frac{1}{2}(\lambda_Q + (\epsilon + \xi_+)\lambda_2)\rho U\;.
\end{align}
The main term should be $\xi_+ \rho I$ and $\xi_\times \rho I$ owing to the intensity-to-polarization leakage $\rho$ at the HWP and the polarization asymmetry $\xi_+$ and $\xi_\times$ in the optics between the HWP and detector.
The 4f-modulated unpolarized components arising from the $Q/U\rightarrow I$ leakage, $\gamma_Q$ and $\gamma_U$, are calculated according to:
\begin{align}
d^{(4,0,\mathrm{Re})} &=\gamma_Q (Q + Q_1 + \lambda_1 I) - \gamma_U U \;,\\
d^{(4,0,\mathrm{Im})} &=\gamma_U (Q + Q_1 + \lambda_1 I) + \gamma_Q U \;.
\end{align}
The 4f-modulated polarized components, including the main polarization signals, are calculated according to:
\begin{align}
d^{(4,2,\mathrm{Re})} &=\epsilon (Q + Q_1 + \lambda_1 I) \;,\\
d^{(4,2,\mathrm{Im})} &=\epsilon U \;.
\end{align}
The 4f-modulated polarized components flipped by the polarization asymmetry $\xi_+$ and $\xi_\times$ in the optics between the HWP and detector are calculated according to:
\begin{align}
d^{(4,-2,\mathrm{Re})} &=\xi_+ (Q + Q_1 + \lambda_1 I) - \xi_\times U \;,\\
d^{(4,-2,\mathrm{Im})} &=\xi_\times (Q + Q_1 + \lambda_1 I) + \xi_+ U \;.
\end{align}

In addition to the nonidealities of the optics, we consider the detector nonlinearity, which is modeled as:
\begin{equation}
d' = d + g_1 d^2 + \tau_1 (\partial_t d) d\;,
\end{equation}
where $g_1$ and $\tau_1$ are the nonlinearity parameters of the detector gain and time constant~\citep{Takakura2017JCAP}.
The second term combines two modulated components and produces another modulation. For instance, the
$(4,4,\mathrm{Re})$ component may arise from:
\begin{equation}
\{\cos(2\theta + 2\phi)\}^2 = \frac{1}{2} + \frac{1}{2}\cos(4\theta + 4\phi)\;,
\end{equation}
or
\begin{equation}
\cos2\phi\cos(4\theta + 2\phi) = \frac{1}{2}\cos4\theta + \frac{1}{2}\cos(4\theta + 4\phi)\;.
\end{equation}
Similarly, $(4,4,\mathrm{Im})$, $(2,4,\mathrm{Re})$, and $(2,4,\mathrm{Im})$ components can appear via the nonlinearity.

These $15 + 4$ components are listed in \cref{tab:tod components} and considered in the detector timestream decomposition.
We can consider more components due to nonlinearity: $(0, 4,\mathrm{Re})$, $(0, 4,\mathrm{Im})$,  $(4, -4,\mathrm{Re})$, and  $(4, -4,\mathrm{Im})$. In practice, however, there are only four types of polarization angle for each detector wafer, and these modes are degenerated with other modes owing to aliasing. Therefore, they are not included in the analysis.

\begin{deluxetable*}{lrcl}
\tablecaption{Decomposition of components
in detector timestreams based on the dependence on the HWP angle $\theta$ and detector angle $\phi$.
The notation column gives the coefficients of $\theta$ and $\phi$ and indicates the real or imaginary component of the demodulated data.\label{tab:tod components}}
\tablehead{
\colhead{Notation} & \colhead{Modulation} & \colhead{Signal} & \colhead{Potential source of the component}
}
\startdata
$(0,0,\mathrm{Re})$ & $1$ & $I$ & Intensity signal \\
$(0,2,\mathrm{Re})$ & $\cos(2\phi)$ &  & Polarized emission in the optics between the HWP and detector \\
$(0,2,\mathrm{Im})$ & $-\sin(2\phi)$ &  & Intensity-to-polarization leakage in the optics between the HWP and detector \\
\hline
$(2,0,\mathrm{Re})$ & $\cos(2\theta)$ & $\rho Q$ & Polarization-to-intensity leakage at the HWP \\
$(2,0,\mathrm{Im})$ & $-\sin(2\theta)$ & $\rho U$ & Polarization-to-intensity leakage at the HWP \\
$(2,2,\mathrm{Re})$ & $\cos(2\theta + 2\phi)$ & $\rho I$ & Intensity-to-linear polarization leakage at the HWP \\
$(2,2,\mathrm{Im})$ & $-\sin(2\theta + 2\phi)$ & $s V$ & Circular polarization to linear polarization leakage at the HWP \\
$(2,-2,\mathrm{Re})$ & $\cos(2\theta - 2\phi)$ &  & Polarization asymmetry in the optics between the HWP and detector \\
$(2,-2,\mathrm{Im})$ & $-\sin(2\theta - 2\phi)$ &  & Polarization asymmetry in the optics between the HWP and detector \\
$(2,4,\mathrm{Re})$ & $\cos(2\theta + 4\phi)$ &  & Detector nonlinearity \\
$(2,4,\mathrm{Im})$ & $-\sin(2\theta + 4\phi)$ &  & Detector nonlinearity \\
\hline
$(4,0,\mathrm{Re})$ & $\cos(4\theta)$ &  & Polarization-to-intensity leakage in the optics between the HWP and detector \\
$(4,0,\mathrm{Im})$ & $-\sin(4\theta)$ &  & Polarization-to-intensity leakage in the optics between the HWP and detector \\
$(4,2,\mathrm{Re})$ & $\cos(4\theta + 2\phi)$ & $Q$ & Stokes $Q$ polarization signal \\
$(4,2,\mathrm{Im})$ & $-\sin(4\theta + 2\phi)$ & $U$ & Stokes $U$ polarization signal \\
$(4,-2,\mathrm{Re})$ & $\cos(4\theta - 2\phi)$ &  & Polarization asymmetry in the optics between the HWP and detector \\
$(4,-2,\mathrm{Im})$ & $-\sin(4\theta - 2\phi)$ &  & Polarization asymmetry in the optics between the HWP and detector \\
$(4,4,\mathrm{Re})$ & $\cos(4\theta + 4\phi)$ &  & Detector nonlinearity \\
$(4,4,\mathrm{Im})$ & $-\sin(4\theta + 4\phi)$ &  & Detector nonlinearity \\
\enddata
\end{deluxetable*}

\vspace*{-2\intextsep}
Note that we use measured angles $\theta$ and $\phi$ in the demodulation (\cref{eq:d4f,eq:d2f}) and decomposition (\cref{eq:decomposition}). The calibration error of these angles may rotate the phase between $d^{(a,b,\mathrm{Re})}$ and $d^{(a,b,\mathrm{Im})}$ as discussed in \cref{sec:polangleerror}. In the case of oblique incidence on the HWP, the modulation angle may have a systematic error due to the projection effect, depending on its angle.
This can be expressed by:
\begin{equation}
\theta \approx \tilde{\theta} + \sigma_+ \sin 2 \tilde{\theta} - \sigma_\times \cos 2 \tilde{\theta}\;,
\end{equation}
where
\begin{equation}
\sigma_+=\frac{n_x^2-n_y^2}{4}\quad\text{and}\quad\sigma_\times=\frac{n_x n_y}{2}\;,
\end{equation}
with $\vec{n} = (n_x, n_y, n_z)$ representing the normal vector of the incident direction on the HWP. This effect may cause the mixing of modes as discussed in Appendix~\ref{sec:polleakagesubtraction}.

\section{Leakage subtraction}\label{appendix:leakage}
Using the instrumental model described in Appendix~\ref{sec:components}, we model the measured circular polarization signal $\tilde{d}^{(2,2,\mathrm{Im})}$ as:
\begin{equation}
\tilde{d}^{(2,2,\mathrm{Im})} \approx \epsilon sV
+ \lambda^{I\rightarrow sV} I
+ \lambda^{Q\rightarrow sV} Q
+ \lambda^{U\rightarrow sV} U + \mathrm{const.}
\end{equation}

The intensity leakage may arise from the optics and detector nonlinearity, as expressed by
$\lambda^{I\rightarrow sV} = \lambda^{I\rightarrow sV}_\mathrm{opt} + \lambda^{I\rightarrow sV}_\mathrm{nl}$, where
\begin{align}
\lambda^{I\rightarrow sV}_\mathrm{opt} &= - \frac{1}{2}(\lambda_U + \xi_\times\lambda_2)\rho\lambda_1\:,\\
\lambda^{I\rightarrow sV}_\mathrm{nl} &\approx 2\omega \tau_1 \epsilon (\rho I + Q_\mathrm{h})\;,
\end{align}
and $\omega$ is the angular speed of the HWP. In a later section, we find $\lambda^{I\rightarrow sV}\approx0.1\%$, which should result from $\lambda^{I\rightarrow sV}_\mathrm{nl}$.

The polarization leakage may arise from the optics, detector nonlinearity, and slant incidence according to:
\begin{align}
\lambda^{Q\rightarrow sV} &= \lambda^{Q\rightarrow sV}_\mathrm{opt} + \lambda^{Q\rightarrow sV}_\mathrm{nl} + \lambda^{Q\rightarrow sV}_\mathrm{si}\;,\\
\lambda^{U\rightarrow sV} &= \lambda^{U\rightarrow sV}_\mathrm{opt} + \lambda^{U\rightarrow sV}_\mathrm{nl} + \lambda^{U\rightarrow sV}_\mathrm{si}\;,
\end{align}
where
\begin{align}
\lambda^{Q\rightarrow sV}_\mathrm{opt} &= - \frac{1}{2}(\lambda_U + \xi_\times\lambda_2)\rho\:,\\
\lambda^{U\rightarrow sV}_\mathrm{opt} &= \frac{1}{2}(\lambda_Q + (\epsilon + \xi_+)\lambda_2)\rho\:,\\
\lambda^{Q\rightarrow sV}_\mathrm{nl} &\approx-3\omega\tau_1 \epsilon \rho Q_1\;,\\
\lambda^{U\rightarrow sV}_\mathrm{nl} &\approx-\frac{1}{2}g_1\epsilon \rho Q_1\;,\\
\lambda^{Q\rightarrow sV}_\mathrm{si} &\approx-2\epsilon\sigma_\times,\\
\lambda^{U\rightarrow sV}_\mathrm{si} &\approx-2\epsilon\sigma_+\;.
\end{align}
In a later section, we find $\lambda^{Q\rightarrow sV}\sim 5\%$ and $\lambda^{U\rightarrow sV}\sim 10\%$. These values are much larger than values obtained for the leakage due to the optics and nonlinearity, estimated to be $< 0.1\%$.
We thus consider the slant incidence effect as the cause of this polarization leakage.

We estimate these leakage coefficients and subtract for the leakage systematics using the measured intensity and linear polarization signals:
\begin{align}
\tilde{d}^{(0,0,\mathrm{Re})} & \approx (1 + \gamma_Q \lambda_2) I\;,\\
\tilde{d}^{(4,2,\mathrm{Re})} & \approx \epsilon Q + \lambda^{I\rightarrow Q} I\;,\\
\tilde{d}^{(4,2,\mathrm{Im})} & \approx \epsilon U + \lambda^{I\rightarrow U} I\;.
\end{align}
Here, intensity-to-linear polarization leakage arises from multiple sources, which is expressed as:
\begin{align}
\lambda^{I\rightarrow Q}_\mathrm{opt} & = \epsilon \lambda_1 \;,\\
\lambda^{I\rightarrow U}_\mathrm{opt} & = 0 \;,\\
\lambda^{I\rightarrow Q}_\mathrm{nl} & \approx g_1 \epsilon Q_1 \;,\\
\lambda^{I\rightarrow U}_\mathrm{nl} & \approx 4 \omega \tau_1 \epsilon Q_1\;,\\
\lambda^{I\rightarrow Q}_\mathrm{si} & \approx \epsilon \rho \sigma_+ \;,\\
\lambda^{I\rightarrow U}_\mathrm{si} & \approx \epsilon \rho \sigma_\times\;.
\end{align}
Note that the estimated leakage coefficients may differ slightly from the modeled leakage coefficients:
\begin{align}
\tilde{\lambda}^{I\rightarrow X} &= \lambda^{I\rightarrow X}/(1 + \gamma_Q \lambda_2)\;,\\
\tilde{\lambda}^{Q\rightarrow X} &= \lambda^{Q\rightarrow X}/\epsilon\;,\\
\tilde{\lambda}^{U\rightarrow X} &= \lambda^{U\rightarrow X}/\epsilon\;.
\end{align}
We detail the analysis methods and results in the following sections.

\subsection{Intensity leakage}\label{appendix:intensity leakage}
The intensity signal is the component extracted as $\tilde{d}^{(0,0,\mathrm{Re})}$; i.e., the unpolarized component not modulated by the HWP.
In addition to the intensity signal from the sky,
this component may contain fluctuations of the focal plane temperature and variations in the ground emission detected through the telescope's far sidelobe, as the mountain Cerro Toco lies
to the east of the telescope and
there are containers around the telescope.

In the basic model with the nonidealities of the HWP (\cref{sec:detector signal model}), the intensity signal $I$ leaks into the real part of the demodulated timestream of the second harmonic $\mathrm{Re}[\tilde{d}_2(t)]$ as $\rho I$ and does not leak into the imaginary component $\mathrm{Im}[\tilde{d}_2(t)]$ that contains $sV$.
The main reason for the leakage into the imaginary part is the variation in the detector time constant~\citep{Takakura2017JCAP}.
The time constant of a transition-edge sensor bolometer increases with the optical power.
The time-constant variation\footnote{The effect of the time constant averaged over each observation is corrected in the calibration.} 
introduces a variable phase lag into the HWP signal modulation, causing variable leakage from the real to the imaginary part of the HWP synchronous signal.

We measure the coefficient of the leakage from the intensity component to other components using the correlation between them~\citep{Takakura2017JCAP}.
The leakage coefficient $\lambda_{(0,0,\mathrm{Re})}^{(*)}$ is obtained from the eigenvector of the covariance:
\begin{equation}
\mathrm{C} 
= 
\left[\begin{matrix}
\langle {\Delta\tilde{d}^{(0,0,\mathrm{Re})}}^2\rangle & 
\langle \Delta\tilde{d}^{(0,0,\mathrm{Re})} \Delta\tilde{d}^{(*)}\rangle / \sqrt{2} \\
\langle \Delta\tilde{d}^{(0,0,\mathrm{Re})} \Delta\tilde{d}^{(*)}\rangle / \sqrt{2} &
\langle {\Delta\tilde{d}^{(*)}}^2\rangle/2
\end{matrix}\right]\;,
\end{equation}
where $\langle \cdots \rangle$ means the operation of the weighted average (i.e., $\langle X \rangle = \sum_t w(t) X(t) / \sum_t w(t)$) and $\Delta\tilde{d}^{(*)}(t) = \tilde{d}^{(*)}(t) - \langle \tilde{d}^{(*)} \rangle$.
In this calculation, the azimuthal structure is temporarily subtracted using the polynomial coefficients estimated in \cref{sec:azfit}.
We subtract the leakage by scaling the intensity component as $\tilde{d}^{\prime(*)}(t) = \tilde{d}^{(*)}(t) - \lambda_{(0,0,\mathrm{Re})}^{(*)} \tilde{d}^{(0,0,\mathrm{Re})}(t)$.
We then fit the azimuthal slope again.

\begin{figure}
\centering
\includegraphics[width=\columnwidth]{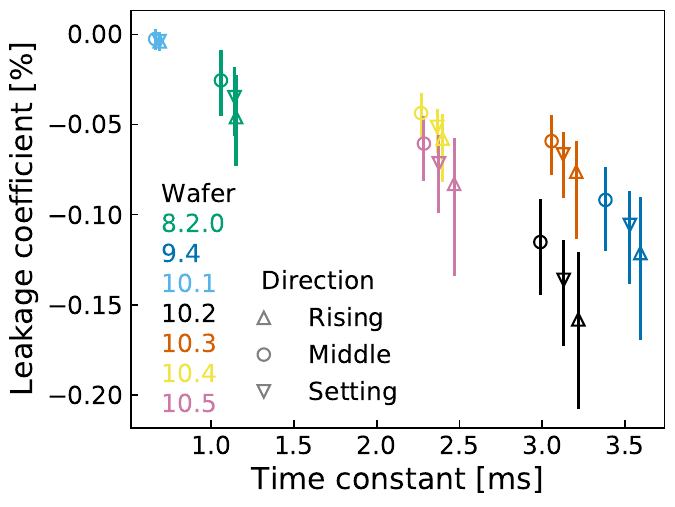}
\caption{Coefficients of the leakage from the intensity $(0,0,\mathrm{Re})$ component to the circular polarization $(2,2,\mathrm{Im})$ component, plotted against the detector time constant. The color and shape of marks denote the wafer and scan direction, respectively. The points show median values and error bars show the observation-by-observation variation. \label{fig:leakage_I}}
\end{figure}

\Cref{fig:leakage_I} shows the estimated coefficients of leakage to the circular polarization $(2,2,\mathrm{Im})$ component.
There is an appreciable dependence on the wafer in that the coefficient is almost zero for wafer 10.1 and approximately $-0.1$\% for the other wafers.
Wafer 10.1 has the largest saturation power and the shortest time constant, and thus the smallest nonlinearity.
For the other wafers, the effects of time constant nonlinearity are observed, with the impact increasing in the slower wafers.
In addition, the detector nonlinearity depends on the optical loading (i.e., the observing elevation and the weather conditions), which causes the variation in the leakage%
\footnote{The variation in the HWP synchronous signal depending on the optical loading also affects the leakage coefficient.}.

After subtraction of this leakage, the variation in the slope $p_{1,\mathrm{obs}}^{(2,2,\mathrm{Im})}$ decreases in all wafers except wafer 10.1.
In particular, the standard deviation decreases by approximately 90\% for wafer 10.2.
Moreover, the coefficient of correlation between $p_{1,\mathrm{obs}}^{(0,0,\mathrm{Re})}$ and $p_{1,\mathrm{obs}}^{(2,2,\mathrm{Im})}$ decreases to $<7\%$ for all wafers.
In addition, the coefficient of correlation between  $p_{1,\mathrm{obs}}^{(2,2,\mathrm{Im})}$ and the leakage coefficient of this intensity subtraction is $<10\%$.
Therefore, the systematic error due to the leakage of the intensity signal is well subtracted, and the residual should have a subdominant effect.

\subsection{Polarization leakage}\label{sec:polleakagesubtraction}
We next consider the leakage of the linear polarization signals, the $\tilde{d}^{(4,2,\mathrm{Re})}$ and $\tilde{d}^{(4,2,\mathrm{Im})}$ components.
The source of these signals should be the instrumental polarization.
The vertical polarization is created by the emission from the primary mirror in the off-axis optics~\citep{Takakura2017JCAP}.
The far sidelobe of the telescope may be polarized and create a polarized ground-synchronous signal.
The contribution of the polarized ice clouds is mitigated by removing scans with polarized bursts during the data selection in \cref{sec:data selection}.

\begin{figure}
    \centering
    \includegraphics[width=0.8 \columnwidth]{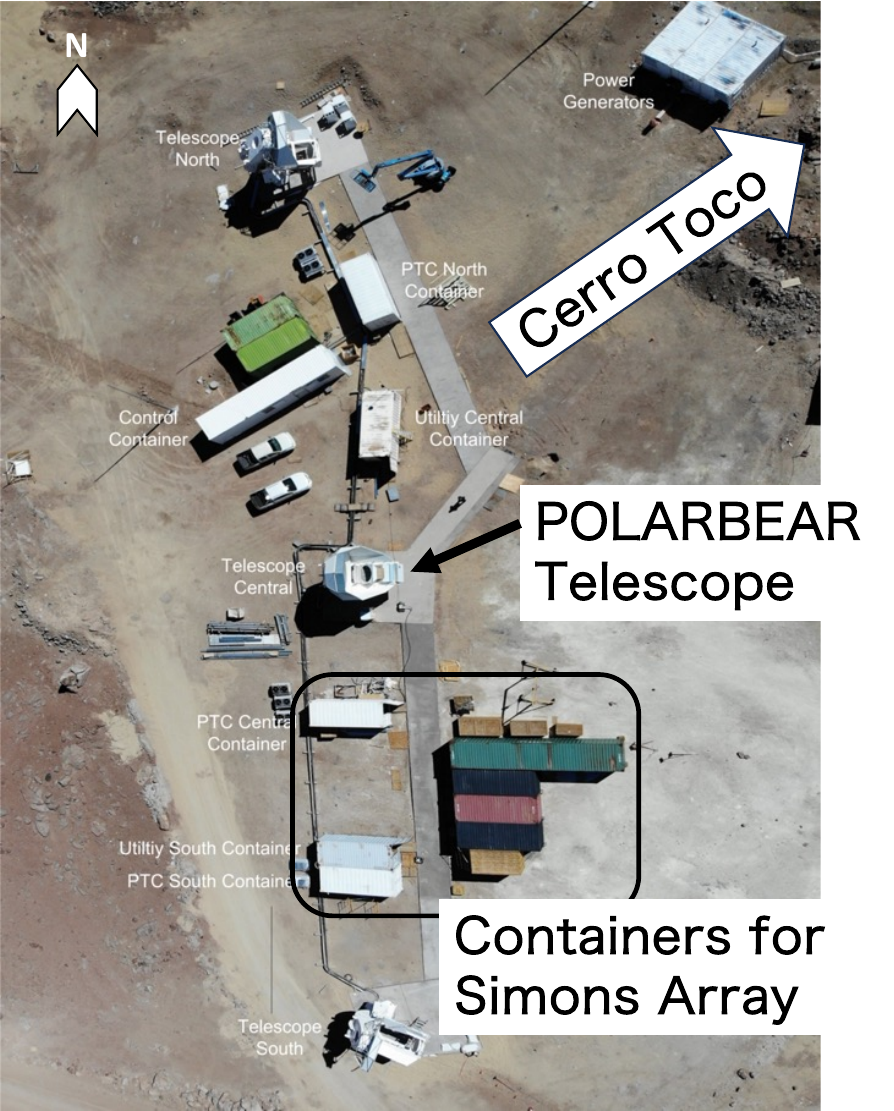}
    \caption{Aerial view of the \textsc{Polarbear}/Simons Array site.
    There are containers for the Simons Array to the south and southeast of the POLARBEAR telescope.
    A mountain, Cerro Toco, lies east of the telescope.}
    \label{fig:google_map}
\end{figure}

As shown in \cref{fig:iqu variation}, we find a linear polarization signal with a non-zero azimuthal slope in both Stokes $Q$ and $U$.
In particular, the rising observations in the fifth season have large slopes of several $\mathrm{mK}/\mathrm{rad}$.
In this scan direction, we observe the lower sky close to the mountain on the east side of the telescope.
Furthermore, the change in the fifth season might be due to the arrival of new containers for deployment of the Simons Array (see \Cref{fig:google_map}).

In the basic model with the nonidealities of the HWP (\cref{sec:detector signal model}), the linear polarization signals $Q$ and $U$ leak into the imaginary part of the demodulated timestream of the second harmonic $\mathrm{Im}[\tilde{d}_2(t)]$ as $\rho(-Q\sin2\phi+U\cos2\phi)$.
Owing to its dependency on the detector polarization angle $\phi$, this leakage should be accounted for separately in the detector averaging in \cref{sec:caltod}, assuming the detector calibrations are accurate.

Another factor to consider is the effect of slant incidence on the HWP.
In the \textsc{Polarbear} telescope, the HWP is placed at the prime focus, oriented so that its surface faces perpendicular to the main optical path.
However, stray light creating the far sidelobe may arrive at a steep incident angle.
In that case, the modulation of polarization signals is distorted, and there is thus leakage from the fourth harmonic to the second and sixth harmonics. This leakage is polarized and not separated by the detector averaging.

We estimate the coefficients of leakage from $(4,2,\mathrm{Re})$ and $(4,2,\mathrm{Im})$ components to the $(2,2,\mathrm{Im})$ component according to the correlation of the azimuthal slopes.
We obtain the leakage coefficients $\lambda_{d}^{(4,2,\mathrm{Re})\rightarrow(2,2,\mathrm{Im})}$ and $\lambda_{d}^{(4,2,\mathrm{Im})\rightarrow(2,2,\mathrm{Im})}$ for each scan direction $d$ by minimizing:
\begin{equation}
\begin{split}
&\sum_{\mathrm{obs}\in d} w_\mathrm{obs}\times \\
&\left(\Delta p_{1,\mathrm{obs}}^{(2,2,\mathrm{Im})} - \lambda_{d}^{(4,2,\mathrm{Re})\rightarrow(2,2,\mathrm{Im})} \Delta p_{1,\mathrm{obs}}^{(4,2,\mathrm{Re})} \right. \\
& \left. - \lambda_{d}^{(4,2,\mathrm{Im})\rightarrow(2,2,\mathrm{Im})} \Delta p_{1,\mathrm{obs}}^{(4,2,\mathrm{Im})}
\right)^2\;.
\end{split}
\end{equation}
Here, to prevent bias due to the atmospheric circular polarization signal's dependence on the scan direction,
we subtract the average for each scan direction $d$ as follows:
\begin{equation}
\Delta p_{1,\mathrm{obs}}^{(*)} = p_{1,\mathrm{obs}}^{(*)} - 
\frac{\sum_{\mathrm{obs}'\in d} w_{\mathrm{obs}'}\,p_{1,\mathrm{obs}'}^{(*)}}{\sum_{\mathrm{obs}'\in d} w_{\mathrm{obs}'}}\;.
\end{equation}
These slopes have been adjusted for intensity leakage subtraction in all three components: $(4,2,\mathrm{Re})$, $(4,2,\mathrm{Im})$, and $(2,2,\mathrm{Im})$.

The statistical uncertainties in the leakage coefficients are estimated adopting the random sign-flip method, a type of bootstrap method.
For each observation, we randomly assign a factor $f \in \{-1, 1\}$ while ensuring that the weighted average remains zero.
We then multiply $\Delta p_{1,\mathrm{obs}}^{(2,2,\mathrm{Im})}$ by this factor and
perform the fitting for each realization.
We repeat this process for many realizations and take the standard deviation of the results.

In addition, we estimate the systematic error from the potential seasonal variations of the leakage coefficients.
In the above calculation,
we use the same leakage coefficient for each wafer and scan direction assuming that it does not change over the seasons.
In practice, however, the leakage coefficient may have seasonal variations.
We estimate the leakage coefficients $\lambda_{d,y}^{(4,2,\mathrm{Re})\rightarrow(2,2,\mathrm{Im})}$ and $\lambda_{d,y}^{(4,2,\mathrm{Im})\rightarrow(2,2,\mathrm{Im})}$ using the subset of data for each season $y$.
We then estimate the systematic uncertainty for each component such that the reduced chi-squared of the difference from the leakage coefficient becomes unity.
This is expressed by:
\begin{equation}
\sum_{y\in\{3,4,5\}} \frac{\left(\lambda_{d,y}^{(4,2,\mathrm{Re})\rightarrow(2,2,\mathrm{Im})} - \lambda_{d}^{(4,2,\mathrm{Re})\rightarrow(2,2,\mathrm{Im})}\right)^2}{\left(\delta\lambda_{d,y,\mathrm{stat}}^{(4,2,\mathrm{Re})\rightarrow(2,2,\mathrm{Im})}\right)^2 + \left(\delta\lambda_{d,\mathrm{sys}}^{(4,2,\mathrm{Re})\rightarrow(2,2,\mathrm{Im})}\right)^2}
= 1\;.
\end{equation}
Finally, the total uncertainty of $\lambda_{d}^{(4,2,\mathrm{Re})\rightarrow(2,2,\mathrm{Im})}$ is estimated as:
\begin{equation}
\begin{split}
\left(\delta\lambda_{d,\mathrm{tot}}^{(4,2,\mathrm{Re})\rightarrow(2,2,\mathrm{Im})}\right)^2 = 
& \left( \delta\lambda_{d,\mathrm{stat}}^{(4,2,\mathrm{Re})\rightarrow(2,2,\mathrm{Im})}\right)^2 \\
& + \left(\delta\lambda_{d,\mathrm{sys}}^{(4,2,\mathrm{Re})\rightarrow(2,2,\mathrm{Im})}\right)^2\;.
\end{split}
\end{equation}
We perform the same calculation for the $(4,2,\mathrm{Im})$ component.

\begin{figure*}
\centering
\includegraphics[width=\textwidth]{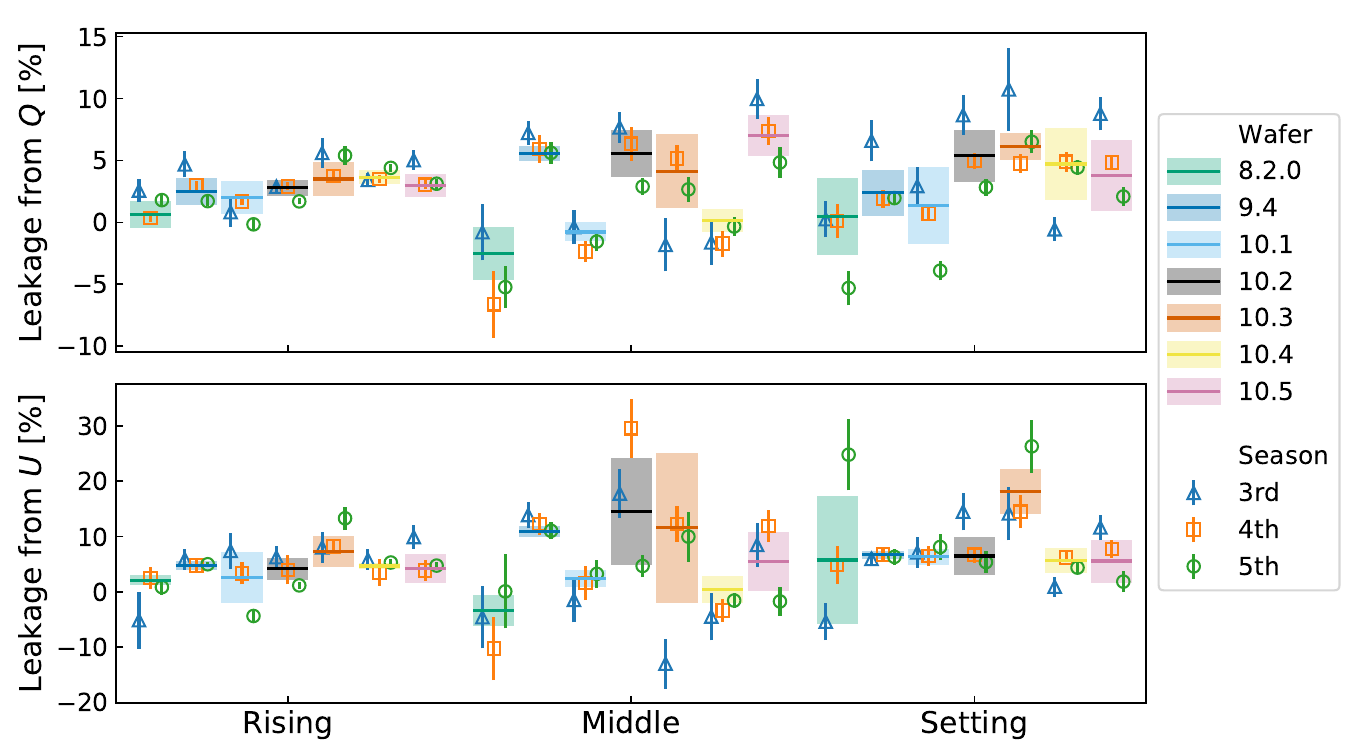}
\caption{Estimated coefficients of the leakage from the linear polarization components $(4,2,\mathrm{Re})$ and $(4,2,\mathrm{Im})$ (corresponding to Stokes $Q$ and $U$) into the circular polarization signal $(2,2,\mathrm{Im})$. Each panel shows the leakage for each wafer and polarization. Each point shows the leakage value and statistical uncertainty for each scan direction and season. A black line shows the value obtained from all data for each scan direction, and a box shows its uncertainty including both the statistical and systematic uncertainties.\label{fig:leakage}}
\end{figure*}

\Cref{fig:leakage} shows the estimated leakage coefficients and their uncertainties.
The amplitudes of the leakage coefficients are $\lesssim 10\%$.

We subtract the leakage for each observation according to:
\begin{equation}
\begin{split}
p_{1,\mathrm{obs}}^{\prime(2,2,\mathrm{Im})}= & p_{1,\mathrm{obs}}^{(2,2,\mathrm{Im})}
- \lambda_{d}^{(4,2,\mathrm{Re})\rightarrow(2,2,\mathrm{Im})} p_{1,\mathrm{obs}}^{(4,2,\mathrm{Re})} \\
& - \lambda_{d}^{(4,2,\mathrm{Im})\rightarrow(2,2,\mathrm{Im})} p_{1,\mathrm{obs}}^{(4,2,\mathrm{Im})}\;.
\end{split}
\end{equation}
This leakage subtraction decreases the standard deviation of slope $p_{1,\mathrm{obs}}^{(2,2,\mathrm{Im})}$ by approximately $10\%$ for wafer 8.2.0 and $80\%$ for wafer 10.4.

\begin{figure}
\centering
\includegraphics[width=\columnwidth]{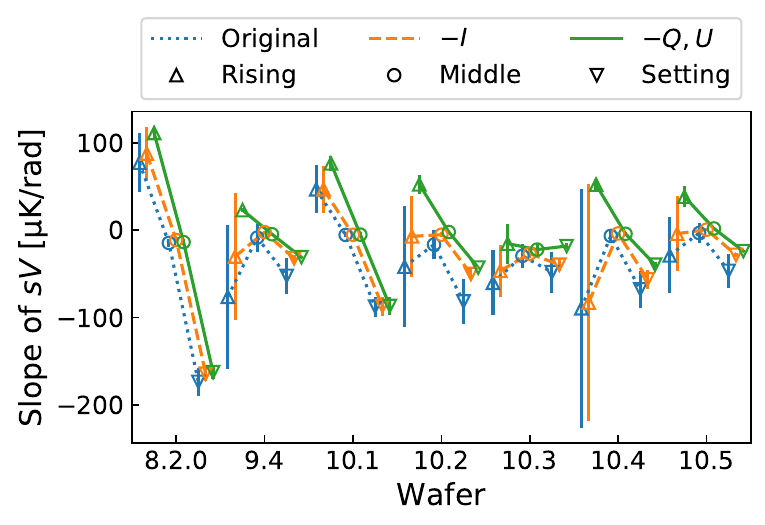}
\caption{
Effect of the intensity and linear polarization subtraction in the circular polarization measurement.
The points show the slope of the circular polarization signals for each wafer and scan direction at different stages of leakage subtraction.
The blue points are values without leakage subtraction.
The orange points are values after subtracting leakage from the intensity signal.
The green points are final values after accounting for leakage from both the intensity and linear polarization.
\label{fig:impact_leak}}
\end{figure}

\Cref{fig:impact_leak} shows the effect of the leakage subtractions.
We find that both intensity and linear polarization subtractions, especially linear polarization subtraction, affect the results of the azimuthal slope.

\bibliography{reference}{}
\bibliographystyle{aasjournal}

\end{document}